\documentclass{article}

\usepackage{arxiv}
\usepackage{authblk}
\usepackage[utf8]{inputenc} 
\usepackage[T1]{fontenc}    
\usepackage{url}            
\usepackage{booktabs}       
\usepackage{amsfonts}       
\usepackage{nicefrac}       
\usepackage{microtype}      
\usepackage{lipsum}
\usepackage[square,numbers]{natbib}

\usepackage{amssymb}
\usepackage{amsthm}
\usepackage{amsmath, esint}
\usepackage{mathtools}
\usepackage{mathrsfs}
\usepackage[ruled, lined, linesnumbered, longend]{algorithm2e}
\usepackage{array}
\usepackage{booktabs}
\usepackage{multirow}
\usepackage{listings}
\usepackage{tabu}
\usepackage{enumerate}
\usepackage{xcolor}
\usepackage[colorinlistoftodos]{todonotes}
\usepackage[colorlinks=true]{hyperref}
\usepackage{textcomp}

\usepackage{graphicx}
\usepackage{subfig}

\usepackage{makecell}
\usepackage[export]{adjustbox}
\usepackage{caption}
\usetikzlibrary{shapes}
\usepackage[para,online,flushleft]{threeparttable}

\usepackage{orcidlink}

\definecolor{myblue}{RGB}{63, 144, 218}
\definecolor{myorange}{RGB}{221, 132, 82}
\definecolor{mygreen}{RGB}{85, 168, 104}
\definecolor{myyellow}{RGB}{255, 169, 14}
\definecolor{myred}{RGB}{189, 31, 1}

\NewDocumentCommand{\mkline}{ O{} O{0.8pt} m }{%
  \raisebox{2pt}{\tikz{\draw[#3, line width = #2, #1] (0,0)--(0.4,0);}}%
}
\newcommand{\mkcross}[2][]{\raisebox{0pt}{\tikz{\draw[#2,line width = 0.8pt, #1] (-.1,-.1) -- (.1,.1) (-.1,.1) -- (.1,-.1);}}}

\newcommand{\mrf}{MuRFiV}
\newcommand{\unet}{U-Net}
\newcommand{\bs}{\boldsymbol}
\newcommand{\bscal}[1]{\boldsymbol{\mathcal{#1}}}

\graphicspath{ {./figs/} }

\DontPrintSemicolon
\definecolor{orcidlogocol}{HTML}{A6CE39}

\title{A Multi-Resolution Finite-Volume Inspired Deep Learning Framework for Spatiotemporal Dynamics Prediction}

\author[1]{Xin-Yang Liu\orcidlink{0000-0003-1423-605X}}
\author[1, 2]{Xiantao Fan\orcidlink{0000-0002-0977-0330}}
\author[1,2,*]{Jian-Xun Wang\orcidlink{0000-0002-9030-1733}}
\affil[1]{Department of Aerospace and Mechanical Engineering, University of Notre Dame, Notre Dame, IN, USA}
\affil[2]{Sibley School of Mechanical and Aerospace Engineering, Cornell University, Ithaca, NY, USA}
\affil[*]{Corresponding author: \texttt{jw2837@cornell.edu}}

\begin{document}
\maketitle

\begin{abstract}
  Predicting complex spatiotemporal dynamics in physical processes often demands computationally expensive numerical methods or data-driven neural networks that suffer from high training costs, error accumulation, and limited generalizability to unseen parameters. An effective approach to address these challenges is leveraging physics priors in training neural networks, known as physics-informed deep learning (PiDL).
  In this work, we introduce the Multi-Resolution Finite-Volume-inspired network, \mrf{}, designed to capitalize on the conservative property of finite volume on the global scale and the expressive power of deep learning on the local scale. We demonstrate the effectiveness of \mrf{} on several spatio-temporal systems governed by partial differential equations (PDEs), including Burgers' equation, shallow water equations, and incompressible Navier-Stokes equations. By embedding PDE information into the deep learning architecture, \mrf{} achieves strong long-term prediction accuracy and remains stable over very long autoregressive rollouts, significantly outperforming data-driven neural network baselines. This result highlights the promise of combining multiresolution learning with finite-volume-inspired inductive bias for accurate and robust long-term prediction of complex dynamics.
\end{abstract}


\section{Introduction}

Spatiotemporal dynamics are central to many problems in science and engineering, including shock formation, transport and mixing, turbulence, and large-scale environmental prediction.
Many of these systems are naturally posed as initial-boundary value problems for partial differential equations (PDEs).
Obtaining numerically reliable solutions at useful spatial and temporal resolutions, however, remains computationally demanding.
Stability restrictions such as the Courant--Friedrichs--Lewy condition, the need to resolve disparate scales, and repeated simulations for design, control, or uncertainty quantification can together lead to large wall-clock costs even on modern hardware.
Classical finite-difference, finite-volume, finite-element, and spectral methods offer well-established accuracy and stability properties, but they often require dense meshes and small time steps to control truncation error over long horizons.

Deep-learning-based surrogate models seek to amortize this cost by replacing repeated numerical solves with a training stage followed by fast inference.
Data-driven models have achieved strong predictive accuracy on canonical PDE benchmarks, including convolutional neural network (CNN) models for Cartesian grids~\cite{fukami2019super}, graph neural network (GNN) simulators for irregular meshes~\cite{pfaff2020learning, han2022predicting}, and neural operators for learning mappings between function spaces across discretizations~\cite{li2020fourier, lu2021learning}.
Despite these successes, purely data-driven surrogates are often black-box models that rely on large amounts of labeled trajectory data and can generalize poorly outside the training distribution.
This limitation is especially consequential for autoregressive rollout, where small one-step errors can compound into unstable or physically inconsistent long-term predictions.

Physics-informed deep learning (PiDL) addresses these limitations by incorporating knowledge of governing laws into the learning problem.
Physics-informed neural networks (PINNs)~\cite{raissi2019physics} and physics-constrained surrogate models~\cite{sun2020surrogate} reduce the dependence on labeled data by penalizing violations of governing equations, conservation laws, initial conditions, and boundary conditions.
These approaches improve data efficiency and physical consistency, but several practical challenges remain.
Multi-objective loss balancing can be sensitive to normalization and weighting, residuals from stiff or strongly nonlinear operators can be difficult to optimize, and satisfying residual constraints during training does not by itself guarantee stable long-term autoregressive prediction.

Recent advances in differentiable solvers have enabled another class of hybrid PiDL methods that couple neural networks with numerical operators and train the combined system end to end.
For example, Bar-Sinai et al.~\cite{bar2019learning} learned data-driven discretizations for PDEs, Kochkov et al.~\cite{kochkov2021machine} accelerated fluid simulation by replacing parts of the Navier--Stokes solver with neural networks, and Fan et al.~\cite{fan2024differentiable} used differentiable hybrid neural modeling for fluid--structure interaction.
Other studies have shown that solver-in-the-loop training and embedded numerical operators can improve long-horizon prediction by introducing useful inductive bias~\cite{um2020solver, liu2024multi, wang2024p}.
Beyond finite-difference-style operators, finite-volume and finite-element ideas have also been incorporated into learnable simulation frameworks to improve conservation, geometric flexibility, and operator reuse~\cite{yan2025learnable, actor2024data, ouyang2026noem}.

Although these hybrid methods have substantially improved the accuracy and robustness of neural surrogates, important gaps remain.
In many approaches, the trainable component is still a generic neural architecture whose internal representation is only weakly aligned with the conservative structure of the underlying dynamics.
Moreover, applying embedded numerical operators on coarse meshes can reduce cost but may lose fine-scale information, whereas applying them densely at high resolution can inherit the computational burden and stability limitations of the original solver.
These issues motivate architectures that encode conservation and multiscale structure directly into the prediction mechanism while retaining the flexibility of learnable models.

In this work, we propose the \emph{Multi-Resolution Finite-Volume-inspired network} (\mrf{}), a deep architecture for long-term prediction of spatiotemporal dynamics.
The central idea is to separate the evolution of subdomain averages from the reconstruction of high-resolution local structure.
At the global scale, \mrf{} uses a finite-volume-inspired global model to predict changes in subdomain averages through fluxes across subdomain interfaces.
At the local scale, a shared local model reconstructs intra-subdomain local fluctuations and enforces a zero-mean constraint so that the reconstructed fine-scale field remains consistent with the global average.
The framework can be used as \mrf{}-Eq, where an embedded numerical operator provides physics-based flux information corrected by the global network, or as \mrf{}-NoEq, where the global network learns the interface fluxes directly.

\paragraph{Contributions.}
\begin{itemize}
  \item \textbf{Finite-volume-inspired architecture for stable roll} We design \mrf{} around domain decomposition, flux-based global updates, and local residual reconstruction, embedding conservative structure directly into the prediction architecture. A lightweight local network is shared across subdomains to reconstruct fine-scale local fluctuations, making high-resolution feature prediction parameter efficient.
out.  \item \textbf{High-quality embedded operators through interface-only flux prediction.} In \mrf{}-Eq, the embedded numerical operator is evaluated on the original fine mesh only along subdomain interfaces, where the global network corrects the fluxes used to update subdomain averages. This design preserves high-quality physics-based flux information at the locations that control global evolution, while keeping the cost low by avoiding dense operator evaluation over the full domain.
  \item \textbf{Robust prediction across canonical PDE systems.} Across one- and two-dimensional Burgers' equations, shallow water equations, and incompressible Navier--Stokes equations, \mrf{} achieves accurate long-term autoregressive prediction, improves stability relative to data-driven baselines, and generalizes to unseen physical parameters.
\end{itemize}

Together, these contributions show that combining finite-volume-inspired conservation with multi-resolution learning provides a practical route to accurate, stable, and efficient neural surrogates for spatiotemporal dynamics.

\section{Methodology}

\subsection{Problem Formulation}
We consider modeling spatiotemporal dynamics governed by general conservative PDEs of the following form:

\begin{align}
    \frac{\partial \bs{u}}{\partial t} = \nabla \cdot \bscal{F}(\bs{u}, \nabla \bs{u}, \nabla^2\bs{u}, \cdots, \bs{\lambda}) + \bscal{Q}(\bs{u}),&  & & \bs{x}, t \in \Omega \times [0, T], \bs{\lambda} \in \mathbb{R}^{m}\\
    \bscal{I}\left[\bs{x}, \bs{u}, \nabla_{\bs{x}}^2\bs{u}, \nabla_{\bs{x}}\bs{u}\cdot\bs{u}, \bs{\lambda}\right] = \bs{0},&  & & \bs{x} \in \Omega, t =0,\bs{\lambda} \in \mathbb{R}^{m}\\
	\bscal{B}\left[t, \bs{x}, \bs{u}, \nabla_{\bs{x}}^2\bs{u}, \nabla_{\bs{x}}\bs{u}\cdot\bs{u}, \bs{\lambda}\right] = \bs{0},&  & & \bs{x}, t \in \partial\Omega \times [0, T], \bs{\lambda} \in \mathbb{R}^{m}\label{eq:pde}
\end{align}

where $\bs{u}(\bs{x}, t)$ is the state variable while $\bs{x}$ and $t$ are the space and time coordinates, respectively. $\bscal{F}$ is a non-linear operator parameterized by $\bs{\lambda}$, while the $\bscal{Q}$ represents any other terms that cannot be formulated as divergence (e.g. body force, volumetric source/sink term). 

Traditionally, numerical methods are leveraged to resolve these spatiotemporal dynamics, where the numerical operator $\bscal{S}$ can be formulated as:
\begin{equation}
\begin{split}
    \bs{u}_{t_i} = \bscal{S}(\bs{u}_{t_{i - 1}},\bs{\lambda}),\quad i\in\{1,2,\cdots,N\}\\
\end{split}
\end{equation}
where $\bs{u}_{t_i}$ denotes the state variable at the $i^{th}$ time step ($t_i$). Please note that the time interval ($[t_{i-1}, t_i]$, $\Delta T= t_i -t_{i-1}$) is generally determined by the dynamics, which can be much longer than the actual numerical step size ($\Delta t$, $\Delta t \ll \Delta T$) used by the solver due to numerical stability consideration, and the notation $\bscal{S}$ may represent the composite operator of many numerical time iterations. 
In this work, we are interested in training a neural surrogate model $\bscal{N}$ given the training dataset $\mathbb{S}$ and a validation set $\mathbb{V}$ such that: 
\begin{equation}
\begin{split}
    &\min_{\bs{\theta}} \mathcal{L}(\{\tilde{\bs{u}}_{t_i}\}, \mathbb{V})\\
    \text{where, } &
    \begin{cases}
        \tilde{\bs{u}}_{t_i} = \bscal{N}(\tilde{\bs{u}}_{t_{i - 1}},\bs{\lambda};\bs{\theta})\\
        \cdots\\
        \tilde{\bs{u}}_{t_1} = \bscal{N}(\bs{u}_{t_{0}},\bs{\lambda};\bs{\theta}), \quad\bs{u}_{t_0} \in \mathbb{V}    
    \end{cases}
\end{split}
\end{equation}
where $\bs{\theta}$ represents the trainable parameters of the neural surrogate model, and $\mathcal{L}$ is the loss function. In this work, Mean Squared Error (MSE) was used as the loss function. Once the model is properly trained, the spatiotemporal trajectory can be predicted by rolling out the surrogate model auto-regressively. 

\subsection{Design Overview of \texorpdfstring\mrf{}}
\begin{figure}[!h]
    \centering
    \includegraphics[width=\textwidth]{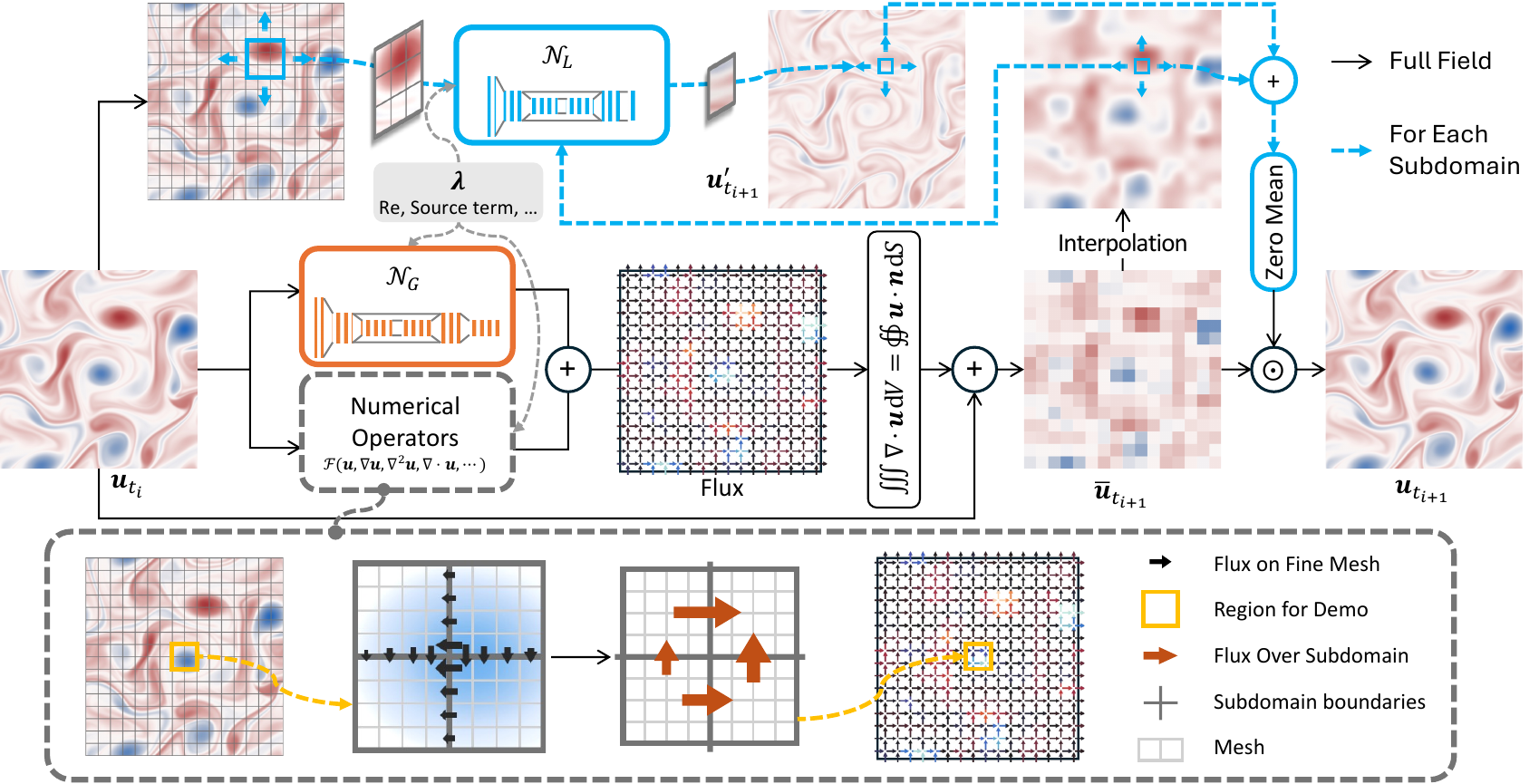}
    \caption{Schematic of \mrf{}. \mrf{} consists of three components: local neural network ($\bscal{N}_L$, shown in the blue box), global neural network ($\bscal{N}_G$, shown in the orange box), and an optional numerical operators component approximating flux ($\bscal{F}$, shown in dashed black box). The local network handles local feature mapping and the same local network is shared over the entire computational domain. The global network learns to predict the flux (when no numerical operators are used) at subdomain interfaces or to correct the error of such flux (when numerical operators are presented). The lower dashed black box illustrates how the embedded numerical operator works. The fluxes are only calculated along the subdomain boundaries based on the original fine mesh (shown by the white arrows), and the flux across the subdomain equals the summation of these (represented by the dark orange arrows). }
    \label{fig:schem}
\end{figure}

\mrf{} consists of two major trainable components: the global network $\bscal{N}_G$ and the local network $\bscal{N}_L$, together with an optional component containing numerical operators. The detailed structure schematic of \mrf{} is shown in Fig.~\ref{fig:schem}. 
In particular, the global network is designed to capture large-scale dynamics while the local network focuses on learning small-scale features. Domain decomposition is leveraged to achieve multi-scale learning effectively and the entire computational domain is decomposed into connected and disjoint subdomains. In contrast to many previous hybrid neural solvers, where numerical operators work on down-sampled resolution for lower computational cost at the expense of degraded accuracy or missing details, the numerical operators in \mrf{} calculate the average change over each subdomain by evaluating the fluxes over subdomain interfaces based on the results at the original high resolution. In such a way, not only high accuracy is preserved but also it is achieved at a relatively low computational cost thanks to the spatial sparsity of the subdomain interfaces. 

\subsection{Domain Decomposition}
For a $D$-dimensional computational domain defined in a Euclidean space $\Omega = [l_d, h_d]^D$, where $l_d$ and $h_d$ represent the lower and higher boundary coordinates of the $d^{th}$ dimension, respectively. We decompose the computational domain into $N_s = \prod_{d=1}^D n_d$ connected and disjoint subdomains ${\Omega}_{\bs{j}}$, where $\bs{j} = [j_1,\cdots,j_d,\cdots, j_D]$, $\bs{j}\in \bs{J}$, $j_d \in \{1,2,\cdots,n_d\}$ represents the index of the subdomain $\Omega_{\bs{j}}$. In this work, we use uniform decomposition for ease of implementation, where every subdomain is an orthotope (e.g., line segment in 1D, rectangle in 2D, cuboid in 3D, ...) and is identical to the others. 

\subsection{Global Model}
For conservative systems, the Finite Volume Method calculates the integral average of the state variable by applying Gauss's Law to the governing equation (Eq.~\ref{eq:pde}):
\begin{equation}
    \frac{\partial\bar{\bs{u}}_{\bs{j}}}{\partial t} V_{\bs{j}} = \iiint_{\Omega_{\bs{j}}} \frac{\partial \bs{u}}{\partial t}\text{ d}V 
    = \oiint_{\partial \Omega_{\bs{j}}} \bscal{F}(\bs{u}, \nabla \bs{u}, \nabla^2\bs{u}, \cdots,\,\bs{\lambda})\cdot\bs{n} \text{ d}S + \iiint_{\Omega_{\bs{j}}}\bscal{Q}_i(\bs{u})\text{ d} V
    \label{eq:consPDE}
\end{equation}
where $\bar{\bs{u}}_{\bs{j}}$ is the integration average over the subdomain $\Omega_{\bs{j}}$ ($\bar{\bs{u}}_{\bs{j}} = \frac{1}{V_{\bs{j}}}\iiint_{\Omega_{\bs{j}}} \bs{u} \text{ d}V$).  $\bscal{F}$ is the numerical operator calculating flux at each subdomain interface and $\bs{n}$ is the unit normal vector of a subdomain interface with an area of $S$. For a given training set $\mathbb{S}$, all state variables are stored at discretized time steps with a fixed time interval $\Delta T$ (i.e. learning step size), where:
\begin{equation}
    \Delta \bar{{\bs{u}}}_{t_i, \bs{j}} = \bar{\bs{u}}_{t_i, \bs{j}} - \bar{\bs{u}}_{t_{i-1}, \bs{j}} = \int_{t_{i-1}}^{t_{i}} \frac{\partial\bar{\bs{u}}_{\bs{j}}}{\partial t} \text{ d}t +\epsilon
\end{equation}
where $\epsilon$ is the discretization and truncation error of the numerical solver $\bscal{S}$. Since this work aims to train a surrogate model, we assume the solution obtained from numerical solvers is the ground truth ($\epsilon\rightarrow0$). \mrf{} uses a global model $\bscal{M}_G$ to predict the average change: $\widetilde{\Delta \bar{{\bs{u}}}}_{t_i} = \bscal{M}_G(\bs{u}_{t_{i-1}}, \bs{\lambda};\,\bs{\theta}_G)$, where $\bs{\theta}_G$ are the trainable parameters. In \mrf{}, a global neural network $\bscal{N}_G$ together with optional embedded numerical operators $\bscal{S}_e$ consists of the global model $\bscal{M}_G$. 
\subsubsection{Modeling Flux between Subdomain} In \mrf{}, the optional embedded numerical operator ($\bscal{S}_e$) approximates the change of the integral average $ \widetilde{\Delta\bar{{\bs{u}}}}^s_{t_i, \bs{j}}$ using forward Euler scheme based on the previous learning step:
\begin{subequations}
\begin{alignat}{2}
    \widetilde{\Delta \bar{{\bs{u}}}}^s_{t_i, \bs{j}} &\approx \frac{\Delta T}{V_{\bs{j}}} \cdot \left. \frac{\partial\bar{\bs{u}}_{\bs{j}}}{\partial t}\right |_{t_{i-1}}\label{eq:euler}\\
    &= \frac{\Delta T}{V_{\bs{j}}}\cdot \left[\left.\oiint_{\partial \Omega_{\bs{j}}} \bscal{F}(\bs{u}, \nabla \bs{u}, \nabla^2\bs{u}, \cdots,\, \bs{\lambda})\cdot \tilde{\bs{n}} \text{ d}S\right|_{t_{i-1}} + 
    \left.\iiint_{\Omega_{\bs{j}}}\bscal{Q}_i(\bs{u})\text{ d} V\right|_{t_{i-1}}\right]\label{eq:flux}\\
    &= \frac{\Delta T}{V_{\bs{j}}} \cdot \bscal{S}_e \left(\bs{u}\left(\bs{x}, t_{i-1}\right),\, \bs{\lambda}\right)\,,\quad \bs{x}\in \partial \Omega_{\bs{j}}\cup\partial \Omega^A_{\bs{j}}
\end{alignat}\label{eq:gflux}
\end{subequations}
where $\Omega^A_{\bs{j}}$ denotes the adjacent subdomains of $\Omega_{\bs{j}}$. The only error introduced in Eq.~\ref{eq:gflux} is due to the large time step ($\Delta T$) used in the Euler discretization (Eq.~\ref{eq:euler}) compared to the ground truth solver $\bscal{S}$. While the spatial discretization used in Eq.~\ref{eq:flux} is mathematically equivalent to $\bscal{S}$ attributed to applying Gauss's Law on the fine mesh. Meanwhile, such operators $\bscal{S}_e$ are only applied to the boundaries of each subdomain. By leveraging Gauss's Law, \mrf{}'s numerical operator $\bscal{S}_e$ can achieve high accuracy at a significantly lower computational cost. 

\subsubsection{Global Neural Network}
When \mrf{} is equipped with the numerical operators $\bscal{S}_e$ (\mrf{}-Eq), a global neural network $\bscal{N}_G$ is designed to correct the error introduced by Eq.~\ref{eq:euler} due to large time step size $\Delta T$:
\begin{equation}
    \bs{F}^{ \mathcal{N}}_c(\bs{x}, t_i) = \bscal{N}_G(\bs{u}_{t_{i-1}},\, \bs{\lambda};\,\bs{\theta}_G)\,,\quad\bs{x} \in \partial \Omega^b, 
\end{equation}
where $\bs{F}^{ \mathcal{N}}_c(\bs{x}, t_i)$ represents the correction of the integration of flux over time $[t_{i-1}, t_i]$ predicted by the global neural network, while $\partial \Omega^b$ denotes the interfaces of all subdomains. Then the change of the integral average $\Delta \bar{{\bs{u}}}_{t_i, \bs{j}}$ for each subdomain $\bs{j}$ is approximated by $\widetilde{\Delta \bar{{\bs{u}}}}_{t_i, \bs{j}}$ where:
\begin{equation}
     \widetilde{\Delta \bar{{\bs{u}}}}_{t_i, \bs{j}} =
     \bscal{M}_G(\bs{u}_{t_{i-1}}, \bs{\lambda};\,\bs{\theta}_G) =
     \frac{\Delta T}{V_{\bs{j}}} \bscal{S}_e \left(\bs{u}\left(\bs{x}, t_{i-1}\right),\,\bs{\lambda}\right) + \sum_{\bs{x}\in\partial \Omega_{\bs{j}}} \bs{F}^{ \mathcal{N}}_c(\bs{x}, t_i)\cdot \bs{n}
\end{equation}
When no equations are leveraged (\mrf{}-NoEq), the Global Network $\bscal{N}_G$ will directly predict the fluxes ($\bs{F}_{\bscal{N}}$) instead of a correction of the fluxes. Thus the integral average change for each subdomain $\bs{j}$ is calculated by:
\begin{equation}
     \widetilde{\Delta\bar{{\bs{u}}}}_{t_i, \bs{j}} = 
     \bscal{M}_G(\bs{u}_{t_{i-1}}, \bs{\lambda};\,\bs{\theta}_G) = 
     \sum_{\bs{x}\in\partial \Omega_{\bs{j}}} \bs{F}^{ \mathcal{N}}(\bs{x}, t_i)\cdot \bs{n}
\end{equation}

\subsection{Local Model}
The local model $\bscal{M}_L$ is designed to predict the variance of the state variable within each subdomain $\bs{u}'_{t_i, \bs{j}}$ and the local model is shared among all the subdomains for efficient use of training data as well as trainable parameters.  $\bs{u}'_{t_i, \bs{j}}$ is defined as:
\begin{equation}
    \bs{u}'(\bs{x}, t_i) = 
    \bs{u}(\bs{x}, t_i) - \bar{\bs{u}}_{t_i, \bs{j}} = 
    \bs{u}(\bs{x}, t_{i-1}) + \Delta \bar{{\bs{u}}}_{t_i, \bs{j}}\,,\quad \bs{x}\in \Omega_{\bs{j}} 
    \label{eq:g_l}
\end{equation}
Based on the definition above, the variable $\bs{u}'_{t_i, \bs{j}}$ has a ``zero-mean'' property for each subdomain: 
\begin{equation}
    \iiint_{\Omega_{\bs{j}}} \bs{u}'_{t_i, \bs{j}} \text{ d}V = 
    \iiint_{\Omega_{\bs{j}}} \bs{u}(\bs{x}, t_i) \text{ d}V - 
    \bar{\bs{u}}_{t_i, \bs{j}}\cdot V_{\bs{j}} 
    = 0 
    \label{eq:zero_mean}
\end{equation}
The local model takes advantage of the ``zero-mean'' property as an extra normalization during the model forward pass when predicting the local variance $\tilde{\bs{u}}'_{t_i, \bs{j}}$, which significantly improves the robustness of \mrf{} in the long-term rollout. 
\begin{subequations}
\begin{alignat}{2}
    \tilde{\bs{u}}'(\bs{x}_{\text{out}}, t_i) &= \bscal{M}_L\left(\bs{u}(\bs{x}_{\text{in}}, t_{i-1}), \bs{\lambda};\bs{\theta}_L\right) \\
    &= \tilde{\bs{u}}'^{ \mathcal{N}}(\bs{x}_{\text{out}}, t_i) - \frac{1}{V_{\bs{j}}} \iiint_{\Omega_{\bs{j}}} \tilde{\bs{u}}'^{ \mathcal{N}}(\bs{x}_{\text{out}}, t_i)\text{ d}V\label{eq:localnn_zm_c}
    \\
    \text{where}\quad \tilde{\bs{u}}'^{ \mathcal{N}}(\bs{x}_{\text{out}}, t_i) &= \bscal{N}_L\left(\bs{u}(\bs{x}_{\text{in}}, t_{i-1}), \bs{\lambda};\bs{\theta}_L\right) + \bscal{I}n(\widetilde{\Delta\bar{{\bs{u}}}}_{t_i, \bs{j}})
    \,,\; \bs{x}_{\text{in}}\in \Omega_{\bs{j}}\cup\Omega^A_{\bs{j}}\,,\; \bs{x}_{\text{out}}\in \Omega_{\bs{j}}
\end{alignat}
\end{subequations}
The $\bscal{N}_L$ denotes the local neural network, with trainable parameters $\bs{\theta}_L$, and $\bscal{I}n$ is a deterministic interpolation function (e.g. nearest, linear, cubic), which up-samples the predicted average change over subdomain $\Omega_{\bs{j}}$ to the mesh resolution. When the local neural network is predicting $\tilde{\bs{u}}'_{}$ at a discretized mesh grid, the integral mean in Eq.~\ref{eq:localnn_zm_c} will degrade to algebraic average. For each subdomain, both the historical information ($\tilde{\bs{u}}'_{t_{i-1},\bs{j}}$) and its adjacent subdomain information ($\tilde{\bs{u}}'(\bs{x}, t_{i-1})$, $\bs{x}\in\Omega^A_{\bs{j}}$) will pose influence on the local variance, thus the local model takes the input of both the query subdomain but also its adjacent subdomains $\Omega^A_{\bs{j}}$. 
In one-dimensional cases, $\Omega_{\bs{j}}^A$ is defined as $\{\Omega_{j_1-1}, \Omega_{j_1+1}\}$, while for two-dimensional cases, $\Omega_{\bs{j}}^A$ is given by: 
$\begin{Bmatrix} 
    \Omega_{j_1-1, j_2-1},& \Omega_{j_1-1, j_2} & \Omega_{j_1-1, j_2+1}\\
    \Omega_{j_1, j_2-1},& \Omega_{j_1, j_2} & \Omega_{j_1, j_2+1}\\
    \Omega_{j_1+1, j_2-1},& \Omega_{j_1+1, j_2} & \Omega_{j_1+1, j_2+1}
\end{Bmatrix}$. 

\subsection{Long-term Prediction of Spatiotemporal Dynamics at High Resolution}
Once we have the Global Model and Local Model ready, \mrf{} predicts the spatiotemporal dynamics at the original high resolution in an autoregressive manner:
\begin{equation}
\begin{split}
    \tilde{\bs{u}}(\bs{x}, t_1) &= \bscal{M}_G(\bs{u}_{t_0, \bs{j}},\; \bs{\lambda};\;\bs{\theta}_G) \,+\, \bscal{M}_L(\bs{u}_{t_0,\bs{j}},\;\bs{\lambda};\;\bs{\theta}_L)\\
    \tilde{\bs{u}}(\bs{x}, t_i) &= \bscal{M}_G(\tilde{\bs{u}}_{t_{i-1}, \bs{j}}, \bs{\lambda}; \bs{\theta}_G) + \bscal{M}_L(\tilde{\bs{u}}_{t_{i-1},\bs{j}},\bs{\lambda};\bs{\theta}_L)\,,\; i\geq2
\end{split}
\end{equation}

\section{Results}
\label{sec:results}
In this section, we demonstrate the performance advantages of two \mrf{} variants in predicting long-term spatiotemporal dynamics, comparing them against \unet{} baselines with slightly larger model size. To ensure a fair comparison, all models undergo the same training procedure, consisting of one-step teacher-forcing followed by autoregressive rollout training. Models are evaluated using the checkpoint with the lowest validation loss. Rollout results on training initial conditions and training physical parameters are provided in Appendix~\ref{sec:more_rollout}.

\subsection{One-dimensional Burgers' Equation}
We begin by evaluating \mrf{} on the one-dimensional viscous Burgers' equation with periodic boundary conditions:
\begin{equation}
    \frac{\partial u}{\partial t} = -u\frac{\partial u}{\partial x} + \nu \frac{\partial^2 u}{\partial x^2}\,,\quad x\in[0,2\pi],\,t\in[0,\infty)
\end{equation}
where $\nu$ denotes viscosity. The Reynolds number is defined as $Re = 1/\nu$.
The training dataset comprises 48 trajectories across four Reynolds numbers $\{20, 80, 140, 200\}$, stored on a uniform mesh with $1024$ cells. Each trajectory starts from an initial condition randomly sampled from a 16-dimensional function space and contains 129 temporal snapshots with a time interval $\Delta T = 0.01$ seconds. Further details on dataset generation are provided in Appendix~\ref{appen:datagen_1db}.
We train the three models on the same dataset with 100 epochs of one-step teacher-forcing training followed by 50 epochs of eight-setp rollout training. In \mrf{}, we divide the domain into $32$ subdomains of the same size, each with $32$ cells.

\begin{figure}[]
    \centering
    \includegraphics[width=\linewidth]{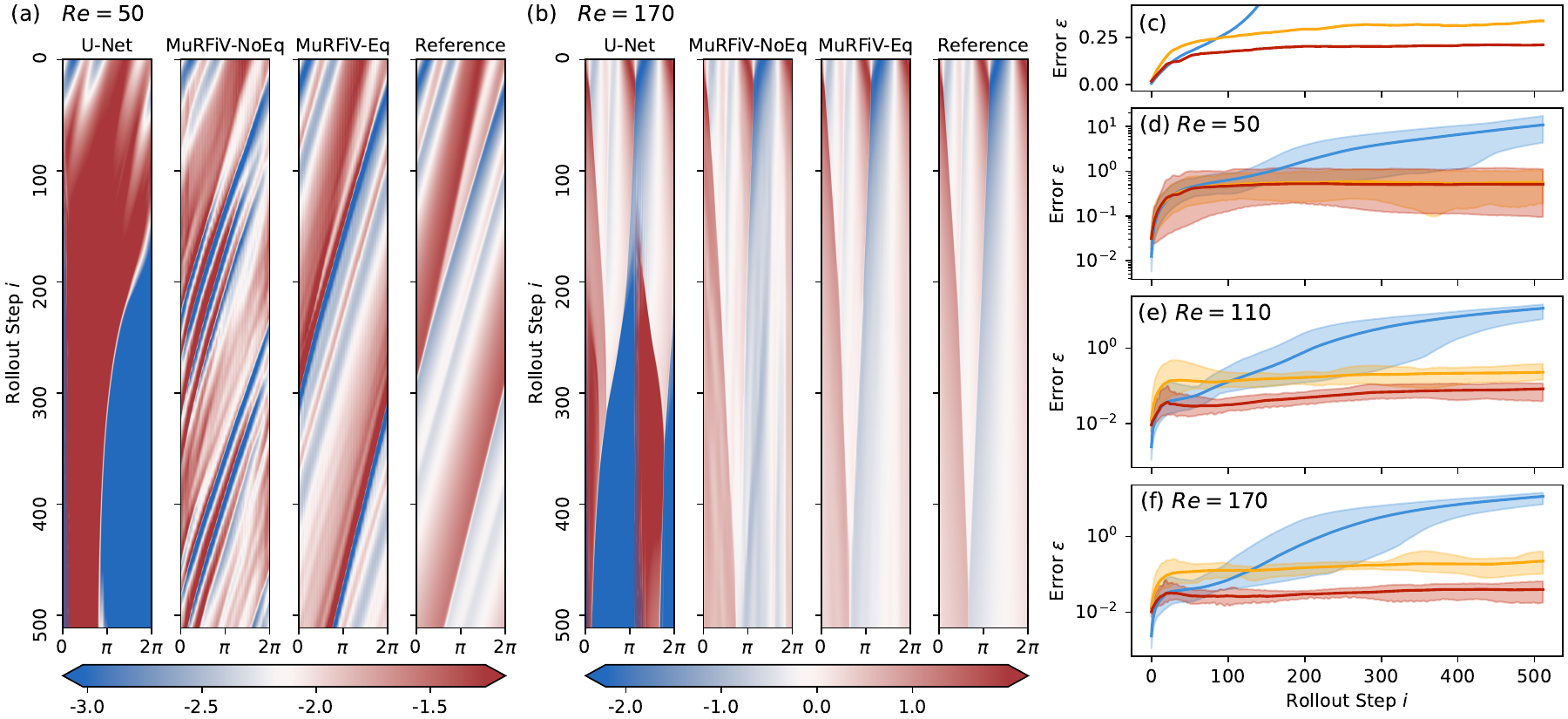}
    \caption{Forecasting performance comparison between \unet{}, \mrf{}-NoEq, \mrf{}-Eq, and the reference data on 1D Burgers' equation. (a) and (b) Predicted velocity ($u$) fields from two representative test cases at different time steps, with Reynolds number $Re = 50$, and $Re=170$, respectively. (c) Change of mean prediction error over time $\epsilon$ averaged over all Reynolds numbers and initial conditions (error in linear scale) and (d-f) averaged error over initial conditions for different test Reynolds numbers (error in log scale). Blue: \unet{} (\mkline{myblue}); Yellow: \mrf{}-NoEq (\mkline{myyellow}); Red: \mrf{}-Eq (\mkline{myred}). Shaded area represents the envelop of error across cases.}
    \label{fig:b1d_test}
\end{figure}

We then investigate the prediction capability of \mrf{} by evaluating its performance on $36$ testing trajectories with unseen Reynolds numbers ($Re\in\{30, 50, 70\}$) and unseen initial conditions drawn from the same distribution. The test dataset therefore probes both long-term rollout accuracy and parameter generalization beyond the training set.

Figure~\ref{fig:b1d_test} presents the forecasting results on this extrapolation task. Visualizations of the predicted velocity field for representative low and high Reynolds numbers are shown in Figure~\ref{fig:b1d_test}(a) and (b). Both \mrf{} variants demonstrate robust generalization, accurately capturing shock formation and propagation despite the shift in physical parameters. In contrast, the \unet{} baseline exhibits significant performance degradation, failing to produce physically consistent predictions in the extrapolation regime.

The quantitative superiority of \mrf{} is reinforced by the error analysis in Figure~\ref{fig:b1d_test}(c-f). Panel (c) reports the average error across all testing cases on a linear scale, while panels (d-f) show the error evolution for specific Reynolds numbers on a log scale. Although \unet{} exhibits a lower error $\epsilon$ than \mrf{} during the initial rollout steps, it quickly accumulates error as autoregressive prediction proceeds. In contrast, both \mrf{} variants maintain stable and accurate predictions over long-term integration, indicating superior robustness against error accumulation. Furthermore, \mrf{}-Eq demonstrates lower prediction error than \mrf{}-NoEq across the unseen Reynolds numbers, underscoring the efficacy of embedding physical operators within the model structure. This enhanced generalizability suggests that the finite-volume backbone and multi-resolution hierarchy enable \mrf{} to learn the underlying physical dynamics rather than merely overfitting to the training data distribution.

\subsection{Two-dimensional Burgers' Equation}
We extend our evaluation to the two-dimensional scalar Burgers' equation with periodic boundary conditions:
\begin{equation}
    \frac{\partial u}{\partial t} = -\left(u\frac{\partial u}{\partial x} +u\frac{\partial u}{\partial y}\right)+ \nu \left(\frac{\partial^2 u}{\partial x^2}+\frac{\partial^2 u}{\partial y^2}\right),\; (x, y)\in[0,2\pi]^2,\;t\in[0,\infty)
    \label{eq:sburgers}
\end{equation}
where $\nu$ represents the viscosity. Similar to the 1D case, the dataset includes trajectories with varying Reynolds numbers ($Re=1/\nu \in \{20, 40, 60, 80\}$), stored on uniform mesh of size $1024\times 1024$. For each Reynolds number, the training trajectories starts from 36 different ICs randomly sampled from a 72-dimensional distributions and contains 129 temporal snapshots with a time interval $\Delta T = 0.01$ seconds. Further details on dataset generation are provided in Appendix \ref{appen:datagen_2db}. The surrogate models are trained on the same dataset with 100 epochs of one-step teacher-forcing training followed by 50 epochs of 8-step rollout training. In \mrf{}, we divide the domain into $32\times32$ sub-domains, each consists with $32\times32$ cells.

\begin{figure}[ht]
    \centering
    \includegraphics[width=\linewidth]{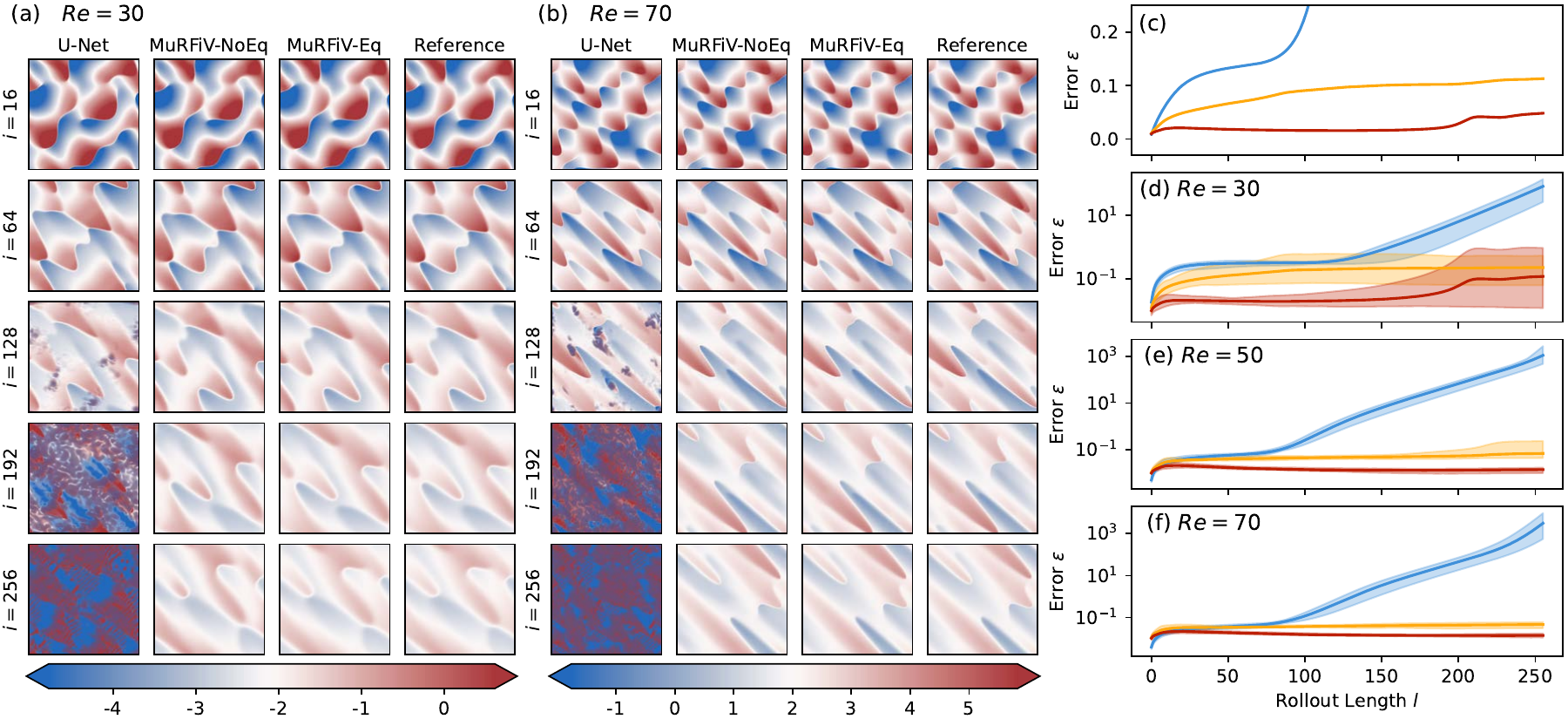}
    \caption{Forecasting performance comparison between \unet{}, \mrf{}-NoEq, \mrf{}-Eq, and the reference data on 2D Burgers' equation. (a) and (b) Predicted velocity ($u$) fields from two representative test cases at different time steps, with Reynolds number $Re = 30$, and $Re=70$, respectively. (c) Change of mean prediction error over time $\epsilon$ averaged over all Reynolds numbers and initial conditions (error in linear scale) and (d-f) averaged error over initial conditions for different test Reynolds numbers (error in log scale). Blue: \unet{} (\mkline{myblue}); Yellow: \mrf{}-NoEq (\mkline{myyellow}); Red: \mrf{}-Eq (\mkline{myred}). Shaded area represents the envelop of error across cases.}
    \label{fig:b2d_Test}
\end{figure}

We next evaluate the prediction capability of \mrf{} on unseen Reynolds numbers ($Re\in\{30, 50, 70\}$) and randomly sampled, unseen initial conditions. The test dataset therefore examines whether the learned dynamics remain accurate when both the physical parameter and the trajectory realization lie outside the training set.

Figure~\ref{fig:b2d_Test} presents the results for this extrapolation task. Visualizations of the predicted scalar field $u$ for representative low ($Re=30$) and high ($Re=70$) Reynolds numbers are shown in Figure~\ref{fig:b2d_Test}(a) and (b). These contours show that \mrf{} robustly handles the shift in physical parameters. Even with unseen Reynolds numbers, \mrf{} correctly predicts the shock speed and sharpness, while preserving the complex interference patterns formed by interacting fronts. The \unet{} baseline, lacking explicit physical constraints, struggles to generalize and exhibits noticeable phase errors together with overly diffused shock structures.

The error analysis in Figure~\ref{fig:b2d_Test}(c-f) confirms the superior generalization of \mrf{}. Panel (c) plots the average error across all test cases on a linear scale, while panels (d-f) show the error evolution for specific Reynolds numbers on a log scale. These error curves show that both \mrf{} variants deliver much more consistent performance across Reynolds numbers, whereas the baseline degrades substantially outside the training regime. Furthermore, \mrf{}-Eq achieves the lowest error throughout most of the rollout, further highlighting the efficacy of embedding physical operators into the model structure. This enhanced generalizability suggests that the embedded finite-volume structure and multi-resolution hierarchy enable \mrf{} to learn the underlying physical operators rather than simply overfitting to the training data distribution.

\subsection{Shallow Water Equation}
\label{sec:sw}
We further assess the efficacy of \mrf{} on the two-dimensional shallow water equations, which govern the flow of a thin layer of fluid under the influence of gravity and viscosity. The system is described by:
\begin{equation}
\begin{split}
    \frac{\partial \eta }{\partial t} + \frac{\partial \eta u}{\partial x}+\frac{\partial \eta v}{\partial y} &= 0\\
    \frac{\partial \eta u}{\partial t} + \frac{\partial}{\partial x}\left(\eta u^2 + \frac{1}{2}g\eta^2\right) & + \frac{\partial (\eta uv)}{\partial y}= \nu\left(\frac{\partial^2 u}{\partial x^2}+\frac{\partial^2 v}{\partial y^2}\right)\\
    \frac{\partial \eta v}{\partial t} + \frac{\partial}{\partial y}\left(\eta v^2 + \frac{1}{2}g\eta^2\right) &+ \frac{\partial (\eta uv)}{\partial x}= \nu\left(\frac{\partial^2 u}{\partial x^2}+\frac{\partial^2 v}{\partial y^2}\right)
\end{split}\,,\,
    (x, y)\in[0,2\pi]^2,\, t\in[0,\infty)
\end{equation}
where $\eta$ denotes the fluid column height, and $(u, v)$ are the velocity components. $g$ is the gravitational acceleration (set to $10$), and $\nu$ represents the viscosity. The training set comprises $80$ different trajectories initialized from random states, with $5$ different viscosities ($1/\nu \in \{1, 2, 4, 8, 16\}$). Further details on dataset generation are provided in Appendix \ref{appen:datagen_swe}. The training data is saved on uniform mesh with $256\times256$ cells, and we divide the mesh equally into $16\times16$ subdomins in \mrf{}.

\begin{figure}
    \centering
    \includegraphics[width=\linewidth]{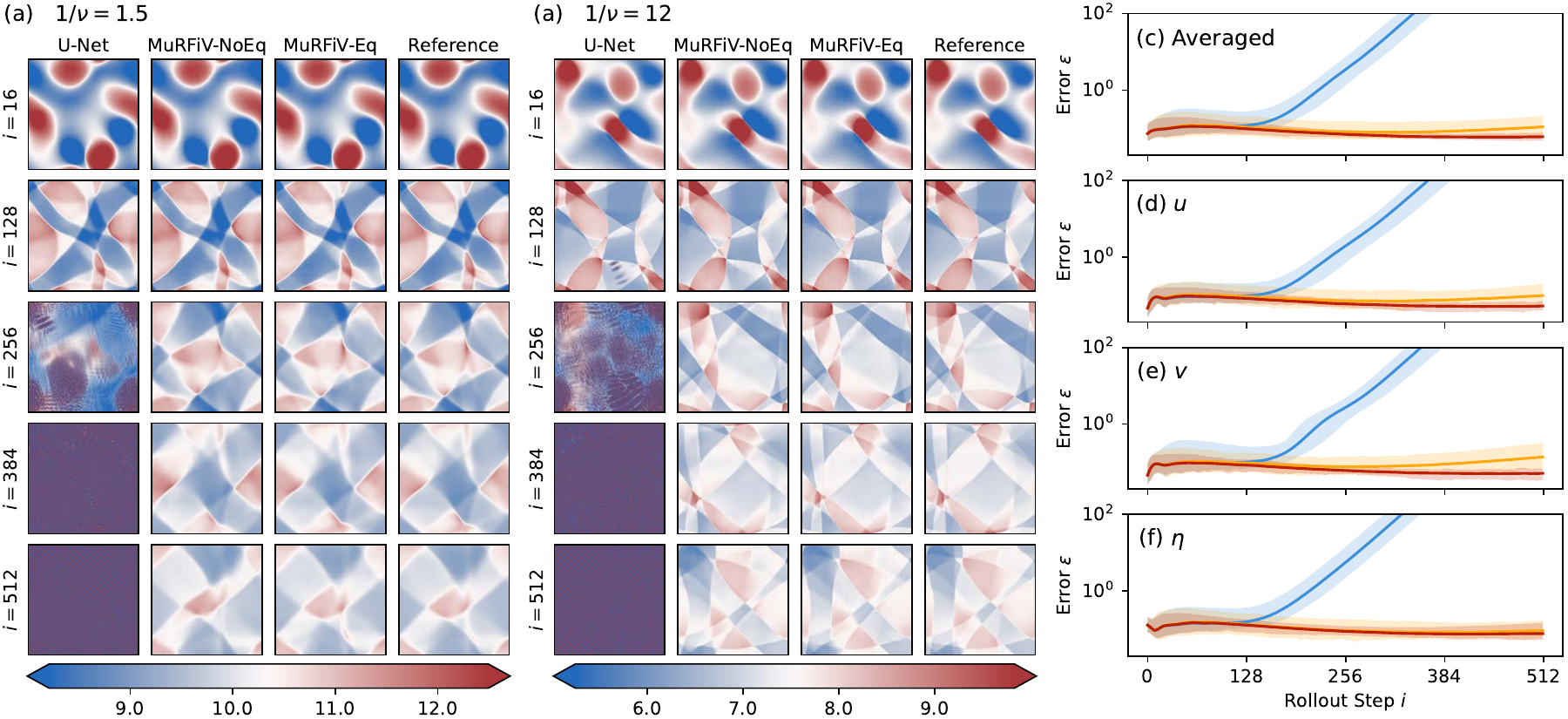}
    \caption{Forecasting performance comparison between \unet{}, \mrf{}-NoEq, \mrf{}-Eq, and the reference data on shallow water equation. (a) and (b) Predicted fluid height ($\eta$) fields from two representative test cases at different time steps, with viscosity $ 1/\nu = 1.5$, and $1/\nu=12$, respectively.  averaged over all Reynolds numbers and initial conditions (error in linear scale) and (d-f) averaged error over initial conditions for different test Reynolds numbers (error in log scale). Blue: \unet{} (\mkline{myblue}); Yellow: \mrf{}-NoEq (\mkline{myyellow}); Red: \mrf{}-Eq (\mkline{myred}). Shaded area represents the envelop of error across cases.}
    \label{fig:sw_test}
\end{figure}

We next evaluate the generalization capability of \mrf{} on $4$ unseen viscosities ($1/\nu \in \{1.5, 3, 6, 12\}$) adn $64$ randomly sampled initial conditions. The test dataset therefore examines whether the learned dynamics remain reliable when the viscous regime changes beyond the training set.

Figure~\ref{fig:sw_test} summarizes the forecasting performance in this extrapolation setting. Panels (a) and (b) visualize representative test cases with the lowest and highest viscosities, respectively. The results indicate that \mrf{} generalizes well to new physical parameters, maintaining physical consistency in wave propagation speeds, amplitudes, and interaction patterns. In contrast, \unet{} exhibits significant error growth and unphysical wave damping when applied to out-of-distribution cases.

The error analysis in Figure~\ref{fig:sw_test}(c-f) confirms the robustness of \mrf{}. Panel (c) presents the aggregate error on a linear scale, while panels (d-f) depict the error evolution for specific viscosities on a log scale. These plots show that \mrf{} outperforms the baseline by a large margin in this challenging extrapolation setting. Notably, \mrf{}-Eq maintains the lowest error profile throughout most of the rollout, further validating that the embedded conservation structure improves long-term stability under unseen viscosities. Similar to the Burgers' cases, the results suggest that the physics-informed structural bias enhances the model's ability to extrapolate to unseen regimes rather than simply memorizing the training distribution.

\subsection{Incompressible Navier-Stokes Equation - Kolmogorov Flow}
Finally, we evaluate \mrf{} on the 2D incompressible Navier-Stokes equations configured for Kolmogorov flow, a classical benchmark for turbulence modeling. The governing equations are:
\begin{equation}
\begin{split}
    & \frac{\partial \bs{u}}{\partial t} = -\left(\bs{u}\cdot\nabla \right)\bs{u} + \nu\nabla^2\bs{u}-\nabla p + \bs{F}_s\\
    & \nabla\cdot \bs{u} = 0
\end{split}\quad \bs{x}\in[0,2\pi]^2,\, t\in[0,\infty)
\end{equation}
where $\bs{u}$ represents the velocity vector, $p$ is the pressure field, and $\nu$ denotes the kinematic viscosity. The flow is driven by a sinusoidal external force $\bs{F}_s = \sin(k_f y)\hat{\bs{x}}$ with forcing wavenumber $k_f=4$. This setup generates a chaotic flow field characterized by a cascade of energy from large to small scales. The Reynolds number is defined as $Re = 1/\nu$. The training dataset consists of $80$ trajectories of fully developed flow across $5$ Reynolds numbers ($Re\in\{128, 256, 512, 1024, 2048\}$), each with $129$ time steps and saved on a $256\times256$ mesh. More details on the dataset generation can be found in Appendix~\ref{appen:datagen_ns}. In \mrf{}, we equally split the $256\times256$ mesh into $16\times16$ subdomains, each with a $16\times16$ mesh.

\begin{figure}[hp]
    \centering
    \includegraphics[width=\linewidth]{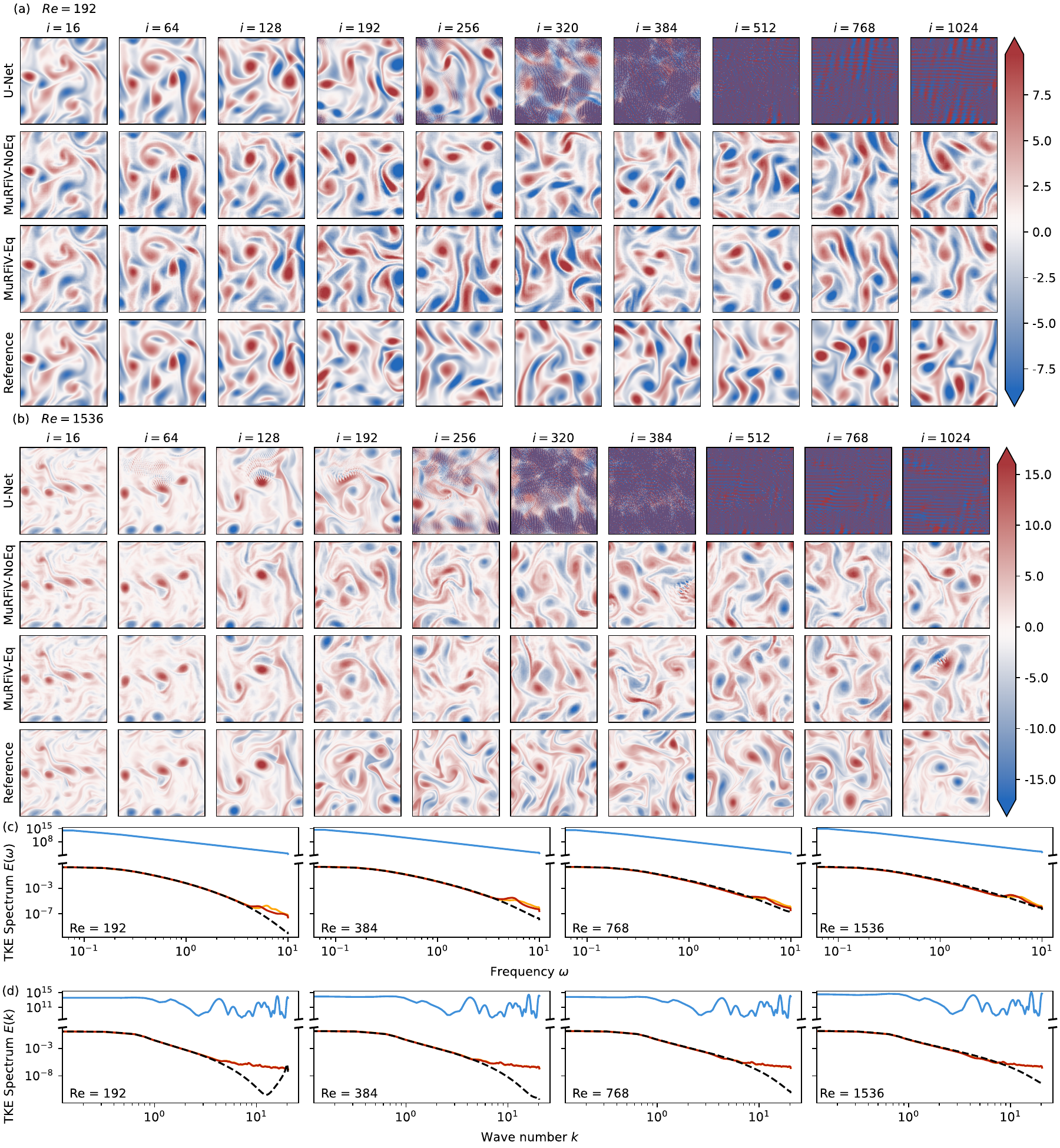}
    \caption{Forecasting performance comparison between \unet{}, \mrf{}-NoEq, \mrf{}-Eq, and the reference data on Kolmogorov flow. (a - b) Instantaneous vorticity ($\omega$) fields at different time steps, of two testing Reynolds numbers $Re=192$ and $Re=1536$, respectively, starting from random unseen initial conditions. (c - d) turbulence kinetic energy (TKE) spectrum comparison at four testing Reynolds numbers. (c) TKE over frequency. (d) TKE over wavenumber. Blue: \unet{} (\mkline{myblue}); Yellow: \mrf{}-NoEq (\mkline{myyellow}); Red: \mrf{}-Eq (\mkline{myred}).}
    \label{fig:ns_test}
\end{figure}

We next evaluate the generalization capability of \mrf{} on $4$ unseen Reynolds numbers ($Re\in\{192, 384, 768, 1536\}$), starting from $64$ random initial conditions. The test dataset therefore examines whether the learned dynamics can preserve both instantaneous flow structures and spectral statistics under a viscosity shift.

Figure~\ref{fig:ns_test} presents the prediciton results on this extrapolation task. Panels (a) and (b) show instantaneous vorticity fields $\omega$ ($\omega = \partial_x v - \partial_y u$) for representative test cases. These visualizations show that \mrf{} robustly adapts to the change in viscosity, producing physically consistent flow patterns and preserving coherent vortical structures better than the \unet{} baseline.

Given the Kolmogorov flow is a chaotic system, we examine the statistics turbulence kinetic energy spectrum instead of absolute error in this section. The spectrum analysis in Figure~\ref{fig:ns_test}(c) and (d) further highlights the superiority of \mrf{}. These panels compare the TKE spectra of the predictions against the DNS reference. Even in this extrapolation regime, \mrf{} maintains a TKE spectrum that closely follows the DNS reference, whereas the baseline fails to capture the correct energy distribution, especially at higher wavenumbers, leading to inaccurate representation of the turbulent statistics. Among the two variants, \mrf{}-Eq provides the closest match to the target flow statistics, further demonstrating the benefit of embedding physical structure into the model. This result shows that the finite-volume-inspired multiresolution design improves not only instantaneous prediction quality but also the preservation of turbulent flow statistics under unseen Reynolds numbers.

\subsection{Design Insight and Ablation Study}
In this section, we justify the design decision of each \mrf{} component by  gradually adding components to the baseline model until reach the full \mrf{} design. All the ablation experiments are tested on the shallow water equation with out-of-distribution initial conditions and viscosity and all the variants has comparable model size (in terms of trainable parameters) as the baseline model (Please refer to Table \ref{tab:ablation_ms})
\begin{figure}[h]
    \centering
    \includegraphics[width=\linewidth]{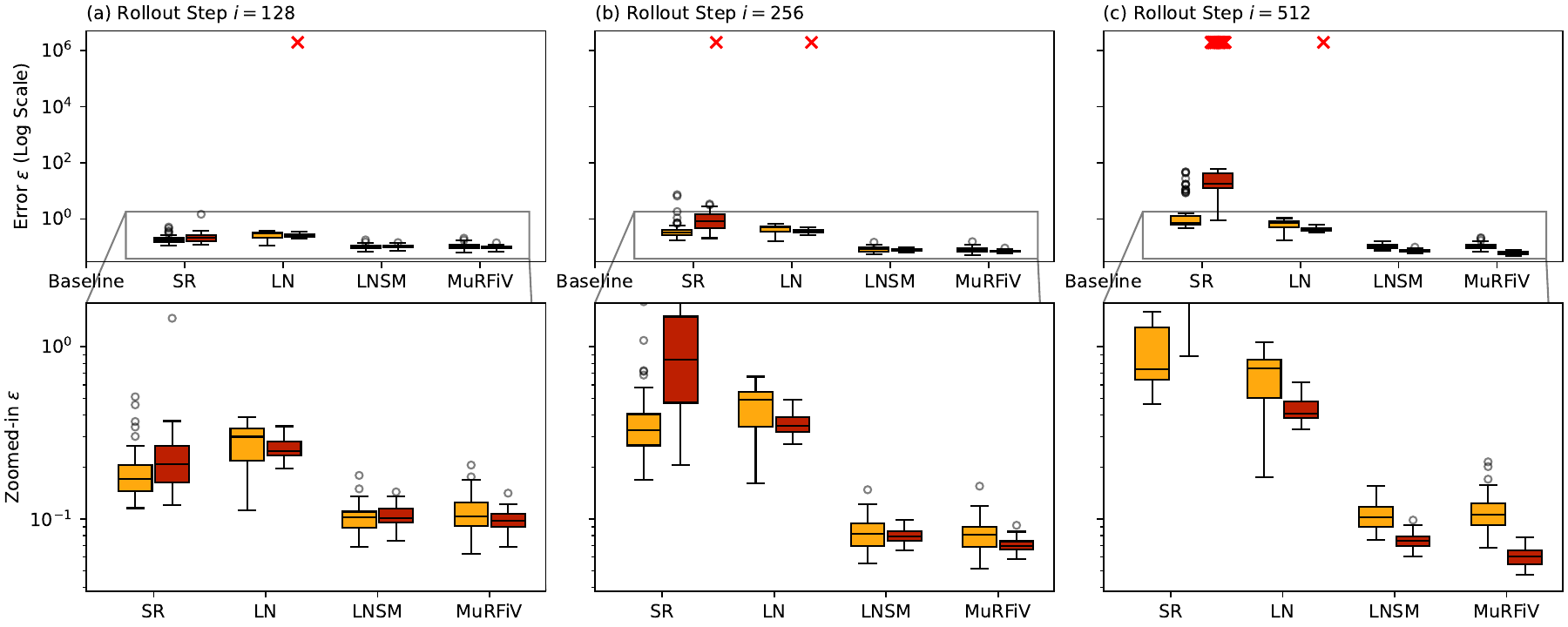}
    \caption{Box plot of prediction error at three different rollout steps (panel a-c, respectively) on $64$ testing trajectories of shallow water equation (same as results shown in Sec \ref{sec:sw}). First row shows the overall view of all variants, while the zoom-in views are shown in the second row. Blue boxes represent the error box of the U-Net baseline, while red and yellow boxes represent the ablation variants with (-Eq) and without (-NoEq) equations embedded, respectively. SR: Global network coupled with a super-resolution network. LN: replace SR network with local neural network. LNSM: local neural network with mean value subtraction. \mrf{}: The complete \mrf{} design, use flux prediction in global network compared to LNSM. Each single red cross (\mkcross{red}) marks the a not-a-number (NaN) value error in the predicted trajectories ((a)1 NaN value in LN-Eq, (b) 1 NaN in SR-Eq and 1 NaN in LN-Eq. (c) 47 NaN in SR-Eq and 1 NaN in LN-Eq) }
    \label{fig:boxplot}
\end{figure}

\begin{table}[h]
    \centering
    \caption{Absolute prediction error at the last timestep ($i=512$) of 64 testing trajectories in ablation study}
    \begin{tabular}{cccccccccc}
    \toprule
    Error ($\epsilon$)  &  \multicolumn{1}{c}{Baseline} & \multicolumn{2}{c}{SR} & \multicolumn{2}{c}{LN} & \multicolumn{2}{c}{LNSM} & \multicolumn{2}{c}{\mrf{}}\\
    \cmidrule(lr){3-4} \cmidrule(lr){5-6} \cmidrule(lr){7-8} \cmidrule(lr){9-10} 
    (Exclude NaN) & (\unet{}) & Eq & NoEq & Eq & NoEq & Eq & NoEq & Eq & NoEq\\
    \midrule
    Mean  & $2.33\times10^5$ & $2.30\times10^{12}$ & {4.90} & $0.434$ & $0.681$ & $0.0747$ & $0.106$ & 0.0605 & 0.110\\
    Median & $1.52\times10^5$ & $18.0$ & 0.641 & 0.409 & 0.491 & 0.0718 & 0.102 & 0.0575 & 0.0968\\
    \# of NaN & 0 & 47 & 0 & 1 & 0 & 0 & 0 & 0 & 0 \\
    \bottomrule         
    \end{tabular}
    \label{tab:ablation_error}
\end{table}

Figure \ref{fig:boxplot} shows the box plot of rollout error at different timesteps, compared across all the structures covered by the ablation study, while the detailed error at the last prediction step is reported in Table \ref{tab:ablation_error}. In this test, we use the same test set used in the Section \ref{sec:sw}. 
\paragraph{Simple Multi-Resolution Structure Improves long-term prediction}
We start improving the baseline method by introducing a multi-resolution structure, noted as SR. SR also consists of two neural networks, one is $\bscal{N}_G$ which predicts the solution field at next step on a coarse mesh ($16\times16$, in this case). The predicted field on the coarse mesh is then up-sampled via bi-cubic to the original resolution and then corrected by a super-resolution network $\bscal{N}_\text{SR}$. In Figure \ref{fig:boxplot}, the multi-resolution structure (SR) already brings noticeable improvement over the U-Net baseline, especially for long-term rollout predictions. Especially when no numerical operator is embedded (SR-NoEq), we lowered the mean prediction error more than four orders of magnitude evaluated at the $512^\text{th}$ rollout steps (See Table \ref{tab:ablation_error}). However, we do observe many ``blow-up'' cases with equation operators embedded, indicating the instability of this structure when couple with numerical operators. 
\paragraph{Shared Local Network is Better Than a Global Super-Resolution Network} Next, we leverage the domain decomposition idea used in \mrf{} by replacing the super-resolution network $\bscal{N}_\text{SR}$ with local variance predictor $\bscal{N}_L$. This variant is noted as LN in Figure \ref{fig:boxplot}. Instead of naively send the whole field predicted by $\bscal{N}_G$ to $\bscal{N}_\text{SR}$ for super-resolution, we decompose the prediction into subdomains which are handled by $\bscal{N}_L$ individually. As reported in Table \ref{tab:ablation_error}, the domain-decomposition structure alone brings over 7 times lower mean prediction error when no numerical operators is embedded (NoEq) and significantly booted the stability (i.e. fewer NaN values) when numerical operators are embedded (Eq).
\paragraph{Guaranteed Zero-Mean for Local Network Prediction Improves Accuracy} We further enhance the LN structure by enforcing a zero-mean output for each subdomain (noted as LNSM in Figure \ref{fig:boxplot} and Table \ref{tab:ablation_error}). Such design choice is inspired by FVM, where the solution in each mesh cell represents the averaged value within the cell. In practice, we guarantee the zero-mean output via subtracting the mean value from the $\bscal{N}_L$, so that $\bscal{N}_L$ only predicts the variance within each subdomain while $\bscal{N}_G$ predicts the mean value of the subdomain. From a neural network perspective, such zero-mean enforcement also serve as a de-bias normalization layer, which also significantly contributes to the long-term rollout stability. The zero-mean/normalization layer design significantly boosted the accuracy over 5 times and eliminated the `blow-up' case when numerical operators are embedded. 
\paragraph{Predicting Flux Instead of Solutions} Lastly, we make the $\bscal{N}_G$ to predict the flux to form the complete structure of \mrf{}. Instead of directly predicting the mean value of each subdomain, $\bscal{N}_G$ predicts the flux at each subdomain interfaces, which ensures adjacent subdomains satisfy conservation: the flux leave one domain must go into its adjacent domains\footnote{Please note that, this does not simply translate to strictly conservation at subdomain level, because the in and out fluxes for a given subdomain are not necessarily balanced. For equations like incompressible Naiver-Stocks, this structure alone cannot guarantee divergence-free}, while the mean value within the subdomain is then obtained via calculating the net fluxes of the given subdomain. From neural network perspective, such design also serves as a normalization, resembling differential output. The flux prediction structure contribute to another $23\%$ accuracy improvement over the LNSM when numerical operators are embedded, though the improvement for NoEq variants is relatively minor. 

From all the ablation tests, we can verify that all the major design components of \mrf{} improves the long term prediction accuracy, under the same model size.  
\subsection{Compare with More Baselines}
\begin{figure}
    \centering
    \includegraphics[width=\linewidth]{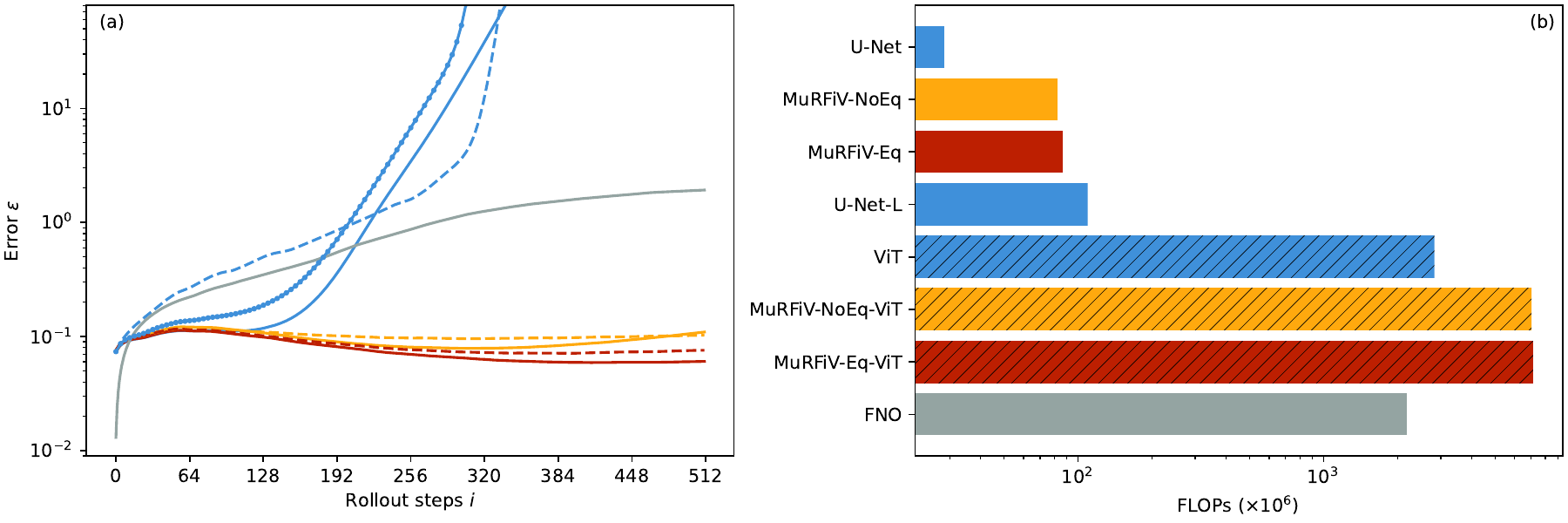}
    \caption{Error and cost comparison. (a) Mean prediction error comparison between multiple baselines at different model rollout steps in the shallow water equation testing trajectories. Grey line marks FNO, blue lines represent classic baselines, yellow lines indicate \mrf{}-NoEq, and red lines represent \mrf{}-Eq. Solid lines represent \unet{}/convolutional networks, while dashed lines represent ViT and \mrf{} ViT variants. Doted line represents the \unet{}-L baseline -  a \unet{} with larger model size. (b) Cost comparison of multiple  baselines in terms of Floating-Point Operations (FLOPs) for one step inference with batch size 1.}
    \label{fig:more_baseline}
\end{figure}

In this section, we compare \mrf{} with other baseline methods, including ViT\cite{dosovitskiy2020image} and FNO\cite{li2020fourier}. We also studied \mrf{} with ViT as the major learning structure. 
\paragraph{Prediction accuracy}
Figure \ref{fig:more_baseline} (a) shows the prediction error averaged on 64 testing trajectories governed by shallow water equation (the same test set as in Section \ref{sec:sw}). The solid lines indicate the convolution neural network based models, while dashed lines represent the ViT based models. Blue lines are classic models like \unet{} and ViT, while the grey line represents FNO. The yellow and red lines represent the \mrf{}-NoEq and \mrf{}-Eq, respectively. The doted blue line represents the error of a larger \unet{}, noted as \unet{}-L. 

On the test set, all the baseline models, including ViT, FNO, and \unet{} of two model sizes, show a significantly error accumulation during long term rollout at different pace. While all the \mrf{} variants managed to maintain the prediction error at a low level, even a slight decrease of the mean absolute error can be observed around step 64-256 due to the intrinsic dissipation property of the dynamics itself. 
Meanwhile, FNO does achieve an impressive lower prediction error for the first few rollout steps on the testing set, and error accumulation rate is relatively slow than other baselines, it still outperformed by all \mrf{} variants by about one order of magnitude lower error at the 512 rollout step. 
\paragraph{Advantage of \mrf{} is independent of neural network types} Interestingly, the increase of the model size does not help improve the rollout accuracy on the testing set. The \unet{}-L even shows a steeper error accumulation curve than the smaller \unet{}. Although ViT shows a slower error accumulation rate around step 128-300, it still fall far behind the ViT based \mrf{} models. Despite the ViT and \unet{} baseline show noticeable error accumulation pattern due to their significant difference in both the model size and model architecture, the corresponding ViT and CNN based \mrf{} variants show a very similar error curve. This indicate the merit of \mrf{} is highly related to the overall architecture design and is independent of specific neural network formulations.

\paragraph{Cost comparison}
Figure \ref{fig:more_baseline}(b) compares the computational cost of all the models studied in this section. Among the baseline methods, FNO has a model size comparable to ViT and is only moderately larger than \mrf{}-Eq and \mrf{}-NoEq, while its one-step inference cost is substantially higher than the CNN-based \mrf{} variants and the \unet{} baselines. Compared with the standard \unet{}, the larger \unet{}-L also requires noticeably higher computational cost, yet this increase does not translate into better rollout accuracy. The ViT-based models are the most expensive in terms of FLOPs, yet the corresponding ViT-based \mrf{} variants still deliver much better rollout accuracy than the plain ViT baseline. The detailed model sizes (numbers of trainable parameters) are listed in Table \ref{tab:ablation_ms}.


\section{Conclusion}
In this work, we introduced \mrf{}, a multi-resolution finite-volume-inspired framework for long-term prediction of spatiotemporal dynamics at the high resolution.
The method separates the evolution of subdomain averages from the reconstruction of intra-subdomain structure: a global model updates subdomain averages through interface fluxes, while a lightweight local model shared across subdomains reconstructs zero-mean local fluctuations in a parameter-efficient manner.
This decomposition embeds the finite-volume view of conservative dynamics into the prediction architecture and provides a direct mechanism for coupling coarse-scale evolution with fine-scale spatial features.

The framework also supports two complementary variants.
In \mrf{}-NoEq, the global network learns interface fluxes directly, showing that the finite-volume-inspired architecture itself provides a strong inductive bias for stable rollout.
In \mrf{}-Eq, an embedded numerical operator evaluates flux information on the original fine mesh only along subdomain interfaces, and the global network corrects these fluxes before updating subdomain averages.
This interface-only use of the embedded operator preserves high-quality physics-based information at the locations that govern subdomain exchange while avoiding dense high-resolution numerical evaluation over the full domain.

Across one- and two-dimensional Burgers' equations, shallow water equations, and incompressible Navier-Stokes equations in the Kolmogorov-flow setting, \mrf{} consistently improves long-term autoregressive prediction relative to blackbox data-driven baselines.
The results show robust generalization to unseen Reynolds numbers, viscosities, and initial conditions, with \mrf{} maintaining shock propagation, wave interactions, coherent vortical structures, and turbulent kinetic-energy spectra more reliably than the baselines.
The ablation study further confirms that the main design choices: multi-resolution coupling, shared local reconstruction, zero-mean local outputs, and flux-based global updates, all contribute to reducing rollout error and improving stability.
Additional comparisons with FNO, ViT, and larger \unet{} baselines indicate that the advantage of \mrf{} comes primarily from the proposed architecture rather than from a specific neural-network backbone or model size.

The present formulation is especially well aligned with conservation-dominated dynamics in which subdomain-average changes can be represented through flux exchange across interfaces.
Future work will extend this idea to more general domain decompositions, complex geometries, non-periodic boundary conditions, and systems requiring additional structure such as strict divergence-free constraints.
It will also be valuable to study adaptive subdomain layouts and more expressive embedded operators for problems where elliptic effects, boundary interactions, or multiscale source terms play a dominant role.
Overall, the results demonstrate that combining finite-volume-inspired global updates with lightweight local reconstruction provides a practical path toward accurate, stable, and efficient neural surrogates for complex spatiotemporal systems.

\section*{Code Availability} All the source codes to reproduce the results in this study will be openly available on GitHub at https://github.com/jx-wang-s-group/MuRFiV upon publication.

\begin{thebibliography}{10}

\bibitem{fukami2019super}
Kai Fukami, Koji Fukagata, and Kunihiko Taira.
\newblock Super-resolution reconstruction of turbulent flows with machine learning.
\newblock {\em Journal of Fluid Mechanics}, 870:106--120, 2019.

\bibitem{pfaff2020learning}
Tobias Pfaff, Meire Fortunato, Alvaro Sanchez-Gonzalez, and Peter Battaglia.
\newblock Learning mesh-based simulation with graph networks.
\newblock In {\em International conference on learning representations}, 2020.

\bibitem{han2022predicting}
Xu~Han, Han Gao, Tobias Pfaff, Jian-Xun Wang, and Li-Ping Liu.
\newblock Predicting physics in mesh-reduced space with temporal attention.
\newblock {\em arXiv preprint arXiv:2201.09113}, 2022.

\bibitem{li2020fourier}
Zongyi Li, Nikola Kovachki, Kamyar Azizzadenesheli, Burigede Liu, Kaushik Bhattacharya, Andrew Stuart, and Anima Anandkumar.
\newblock Fourier neural operator for parametric partial differential equations.
\newblock {\em arXiv preprint arXiv:2010.08895}, 2020.

\bibitem{lu2021learning}
Lu~Lu, Pengzhan Jin, Guofei Pang, Zhongqiang Zhang, and George~Em Karniadakis.
\newblock Learning nonlinear operators via deeponet based on the universal approximation theorem of operators.
\newblock {\em Nature machine intelligence}, 3(3):218--229, 2021.

\bibitem{raissi2019physics}
Maziar Raissi, Paris Perdikaris, and George~E Karniadakis.
\newblock Physics-informed neural networks: A deep learning framework for solving forward and inverse problems involving nonlinear partial differential equations.
\newblock {\em Journal of Computational physics}, 378:686--707, 2019.

\bibitem{sun2020surrogate}
Luning Sun, Han Gao, Shaowu Pan, and Jian-Xun Wang.
\newblock Surrogate modeling for fluid flows based on physics-constrained deep learning without simulation data.
\newblock {\em Computer Methods in Applied Mechanics and Engineering}, 361:112732, 2020.

\bibitem{bar2019learning}
Yohai Bar-Sinai, Stephan Hoyer, Jason Hickey, and Michael~P Brenner.
\newblock Learning data-driven discretizations for partial differential equations.
\newblock {\em Proceedings of the National Academy of Sciences}, 116(31):15344--15349, 2019.

\bibitem{kochkov2021machine}
Dmitrii Kochkov, Jamie~A Smith, Ayya Alieva, Qing Wang, Michael~P Brenner, and Stephan Hoyer.
\newblock Machine learning--accelerated computational fluid dynamics.
\newblock {\em Proceedings of the National Academy of Sciences}, 118(21):e2101784118, 2021.

\bibitem{fan2024differentiable}
Xiantao Fan and Jian-Xun Wang.
\newblock Differentiable hybrid neural modeling for fluid-structure interaction.
\newblock {\em Journal of Computational Physics}, 496:112584, 2024.

\bibitem{um2020solver}
Kiwon Um, Robert Brand, Yun~Raymond Fei, Philipp Holl, and Nils Thuerey.
\newblock Solver-in-the-loop: Learning from differentiable physics to interact with iterative pde-solvers.
\newblock {\em Advances in neural information processing systems}, 33:6111--6122, 2020.

\bibitem{liu2024multi}
Xin-Yang Liu, Min Zhu, Lu~Lu, Hao Sun, and Jian-Xun Wang.
\newblock Multi-resolution partial differential equations preserved learning framework for spatiotemporal dynamics.
\newblock {\em Communications Physics}, 7(1):31, 2024.

\bibitem{wang2024p}
Qi~Wang, Pu~Ren, Hao Zhou, Xin-Yang Liu, Zhiwen Deng, Yi~Zhang, Ruizhi Chengze, Hongsheng Liu, Zidong Wang, Jian-Xun Wang, et~al.
\newblock {P$^2$C$^2$Net}: {PDE}-preserved coarse correction network for efficient prediction of spatiotemporal dynamics.
\newblock {\em Advances in Neural Information Processing Systems}, 37:68897--68925, 2024.

\bibitem{yan2025learnable}
Mengtao Yan, Qi~Wang, Haining Wang, Ruizhi Chengze, Yi~Zhang, Hongsheng Liu, Zidong Wang, Fan Yu, Qi~Qi, and Hao Sun.
\newblock Learnable-differentiable finite volume solver for accelerated simulation of flows.
\newblock In {\em Proceedings of the 31st ACM SIGKDD Conference on Knowledge Discovery and Data Mining V. 2}, pages 3471--3482, 2025.

\bibitem{actor2024data}
Jonas~A Actor, Xiaozhe Hu, Andy Huang, Scott~A Roberts, and Nathaniel Trask.
\newblock Data-driven whitney forms for structure-preserving control volume analysis.
\newblock {\em Journal of Computational Physics}, 496:112520, 2024.

\bibitem{ouyang2026noem}
Weihang Ouyang, Yeonjong Shin, Si-Wei Liu, and Lu~Lu.
\newblock Noem: efficient and scalable finite element method enabled by reusable neural operators.
\newblock {\em Nature Computational Science}, 6(4):417--429, 2026.

\bibitem{dosovitskiy2020image}
Alexey Dosovitskiy, Lucas Beyer, Alexander Kolesnikov, Dirk Weissenborn, Xiaohua Zhai, Thomas Unterthiner, Mostafa Dehghani, Matthias Minderer, Georg Heigold, Sylvain Gelly, et~al.
\newblock An image is worth 16x16 words: Transformers for image recognition at scale.
\newblock {\em arXiv preprint arXiv:2010.11929}, 2020.

\bibitem{jax2018github}
James Bradbury, Roy Frostig, Peter Hawkins, Matthew~James Johnson, Chris Leary, Dougal Maclaurin, George Necula, Adam Paszke, Jake Vander{P}las, Skye Wanderman-{M}ilne, and Qiao Zhang.
\newblock {JAX}: composable transformations of {P}ython+{N}um{P}y programs, 2018.

\end{thebibliography}


\appendix
\setcounter{figure}{0}
\renewcommand{\thefigure}{S\arabic{figure}}
\setcounter{table}{0}
\renewcommand{\thetable}{S\arabic{table}}
\section*{Appendix}
\section{Training and Testing Dataset}
All the training and testing datasets are generated by our finite volume solvers based on Jax\cite{jax2018github}. In this section, we provide more details about how we set up the numerical solver for generating datasets. 
For ease of implementation, all the cases apply periodic boundary conditions at all the boundaries.
\subsection{One-dimensional Burgers' Equation}\label{appen:datagen_1db}
\paragraph{Initial Conditions}
To make the dataset challenging and more representative of the dynamics in Burgers' equation, the training set contains three groups (Group $D$, $D^-$, and $D^+$) of trajectories and each group contains $12$ different trajectories with $4$ different values of Reynolds number $Re \in \{20, 80, 140, 200\}$. Each trajectory within group $D$ starts from different initial conditions $u_{t_0}$, which are randomly sampled from a 16-dimensional random space:
\begin{equation}
    u_{t_0} = \sum_{k\in\mathcal{K}} A_k\sin\left(k x + \varphi_k\right),\quad \mathcal{K} = \{-4, -3, -2, -1, 1, 2, 3, 4\}\label{eq:ic_1db}
\end{equation}
where $A_k$ and $\varphi_k$ are random variables sampled from two uniform distribution $U(0,1)$ and $U(0,2\pi)$, respectively. Since the initial conditions in group $D$ are linear combinations of sinusoidal functions, the average values are all strictly zeros, which will eventually develop into a standing wave regardless of the initial condition. To cover a more comprehensive dynamics pattern of Burgers' equation, we also add groups $D^-$ and $D^+$, where the initial conditions are a simple shift of the ICs in group $D$. Specifically, we generate the ICs in the group $D^-$ by subtracting $2$ from ICs in group $D$ and adding $2$ to form group $D^+$'s ICs. In such a way, the training dataset covers both standing waves and moving waves in both directions in one-dimensional space. Meanwhile, it is still a challenging dataset for neural network training, because merely $12$ different patterns from a $16$-dimensional random space are included in the training set. 

The solution is solved with WENO5 scheme and central difference, with first order forward Euler scheme for time marching. The solver used a time step size $\Delta t = 1\times10^{-4}$s, and we collect the solution every $100$ numerical steps. The training and testing set are sampled from the same initial condition distribution but with different random seed and viscosity. 
\subsection{Two-dimensional Burgers' Equation}\label{appen:datagen_2db}
Two dimensional Burgers' equation case shares almost the same procedure of generating the initial condition as the one-dimensional Burgers'. The only slight difference exists in how we formulate $\bs{u}_{t_0}$:
\begin{equation}
    u_{t_0} = \sum_{k_j\in\mathcal{K}}\sum_{k_i\in\mathcal{K}} A_{k_i, k_j}\sin\left(k_i x + k_j y + \varphi_{k_i,k_j}\right),\quad \mathcal{K} = \{-3, -2, -1, 1, 2, 3\}\label{eq:ic_2db}
\end{equation}
where $A_{k_i, k_j}$ and $\varphi_{k_i, k_j}$ are random variables sampled from two uniform distribution $U(0,1)$ and $U(0,2\pi)$, respectively. In this case, the initial condition is sampled from a 72-dimensional space. All the other parameters like discretization schemes, time step size are identical as the one-dimensional Burgers' equation case.
\subsection{Shallow Water Equation}\label{appen:datagen_swe}
In this case, the initial condition for the fluid column height ($\eta$) is also randomly sampled using the same formula as in the two-dimensional Burgers' equation (Eq.\ref{eq:ic_2db}). The only difference is $\mathcal{K} = \{-2, -1, 0, 1, 2\}$. To ensure we do have a negative fluid column height, the randomly sampled $\eta$ is then linearly shift so that the minium value of the sampled $\eta$ is 1. While the velocity is initialized as zero field. 

We then use first order schemes for both the spatial and temporal derivatives, with a numerical time step size $1\times10^{-6}$s. The data is collected every $5\times10^3$ numerical steps. We discard the first 16 collection steps, to allow the solution field to develop moving waves. 

To resolve the near-discontinuous wave in low viscosity cases while maintain numerical stability in high viscosity cases, the SWE is solved on difference mesh size for different viscosity. The specific mesh size used in preparing the dataset can be found in Table\ref{tab:meshsize}. The training and testing set are then down-sampled from these fine mesh results to a $256\time256$ mesh. 
\begin{table}[]
    \centering
    \caption{Mesh size used in solvers in shallow water equation (top) and Kolmogorov flow (bottom)}
    \begin{tabular}{c|rrrrrrrrr}
    \toprule
        1/$\nu$ &  1 & 1.5 & 2 & 3 & 4 & 6 & 8 & 12 & 16 \\
        \midrule
        Mesh size (Along one axis) & 256 & 256 & 512 & 512 & 1024 & 1024 & 2048 & 2048 & 4096\\ 
        \midrule
        \midrule
        Re &  128 & 192 & 256 & 384 & 512 & 768 & 1024 & 1536 & 2048 \\
        \midrule
        Mesh size (Along one axis) & 256 & 384 & 512 & 768 & 1024 & 1536 & 2048 & 3072 & 4096 \\
    \bottomrule
    \end{tabular}
    \label{tab:meshsize}
\end{table}
\subsection{Navier-Stocks Equation}\label{appen:datagen_ns}
Similarly, we sample the initial condition for velocity components from a distribution resembles Eq.\ref{eq:ic_2db}. Here $\mathcal{K} = \{-1, -1, 0, 1, 1\}$ and $A_{k_i, k_j}\sim U(0, 0.01)$, both x- and y-directional components are sampled from the same distribution, while the pressure field are initialized as zero. We also add a randomly translation in the external force to enrich the dataset. 

We then solve the equation with first order scheme with pressure projection. The numerical solver uses a time step size $\Delta t = 5\times10^{-6}$ and we only collect data after the turbulence is fully developed, at a frequency of every $10^4$ numerical steps. 

Similar to the shallow water equation, we also use different mesh size under different Reynolds numbers (refer to Table.\ref{tab:meshsize} for the exact mesh size used in solvers).  Then the results are all down sampled to the $256\times256$ mesh. 

\section{Model Size}
\begin{table}[h]
    \centering
    \caption{Number of trainable parameters}
    \begin{tabular}{lrrrrr}
    \toprule
    \multirow{2}{*}{Model}  &  \multicolumn{1}{c}{\multirow{2}{*}{\unet{}}} & \multicolumn{2}{c}{\mrf{}-NoEq} & \multicolumn{2}{c}{\mrf{}-Eq} \\
    \cmidrule(lr){3-4} \cmidrule(lr){5-6} 
     & & \multicolumn{1}{c}{$\bscal{N}_G$} & \multicolumn{1}{c}{$\bscal{N}_L$} & \multicolumn{1}{c}{$\bscal{N}_G$} & \multicolumn{1}{c}{$\bscal{N}_L$} \\
     \midrule
    
    \multirow{2}{*}{1D Burgers'} & \multirow{2}{*}{573,019} & \multicolumn{2}{c}{460,654} & \multicolumn{2}{c}{538,446}\\
    \cmidrule(lr){3-4} \cmidrule(lr){5-6} 
     & & 385,811 & 74,843 & 463,603 & 74,843\\
     \\
    
    \multirow{2}{*}{2D Burgers'} & \multirow{2}{*}{3,994,969} & \multicolumn{2}{c}{3,214,873} & \multicolumn{2}{c}{3,755,673}\\
    \cmidrule(lr){3-4} \cmidrule(lr){5-6} 
    & & 2,690,336 & 524,537 & 3,231,136 & 524,537\\
    \\

    \multirow{2}{*}{Shallow Water} & \multirow{2}{*}{3,642,421} & \multicolumn{2}{c}{2,574,557} & \multicolumn{2}{c}{2,729,709}\\
    \cmidrule(lr){3-4} \cmidrule(lr){5-6} 
    & & 2,031,584 & 542,973 & 2,186,736 & 542,973\\
    \\
    
    \multirow{2}{*}{Kolmogrov Flow} & \multirow{2}{*}{8,234,053} & \multicolumn{2}{c}{6,927,485} & \multicolumn{2}{c}{7,494,157}\\
    \cmidrule(lr){3-4} \cmidrule(lr){5-6} 
    & & 5,706,800 & 1,220,685 & 6,273,472 & 1,220,685\\
    
    
    \bottomrule
    \end{tabular}
    \label{tab:my_label}
\end{table}
\begin{table}[h]
    \centering
    \caption{Number of trainable parameters of more baselines and ViT variants}
    \begin{tabular}{cccccc}
    \toprule
    & FNO & U-Net-L & ViT & \mrf{}-NoEq (ViT) & \mrf{}-Eq (ViT)\\
     \midrule 
    \# Trainable Parameters & 3,628,443 & 71,028,253 & 19,206,144 & 15,645,714 & 15,843,218\\   
    \bottomrule
    \end{tabular}
    \label{tab:ablation_ms}
\end{table}

\section{More Rollout Test Results}
\label{sec:more_rollout}
This section summarizes the rollout performance on training initial conditions and training physical parameters. The figures are moved here to keep the main results section focused on generalization to unseen conditions.

\begin{figure}[!h]
    \centering
    \includegraphics[width=\linewidth]{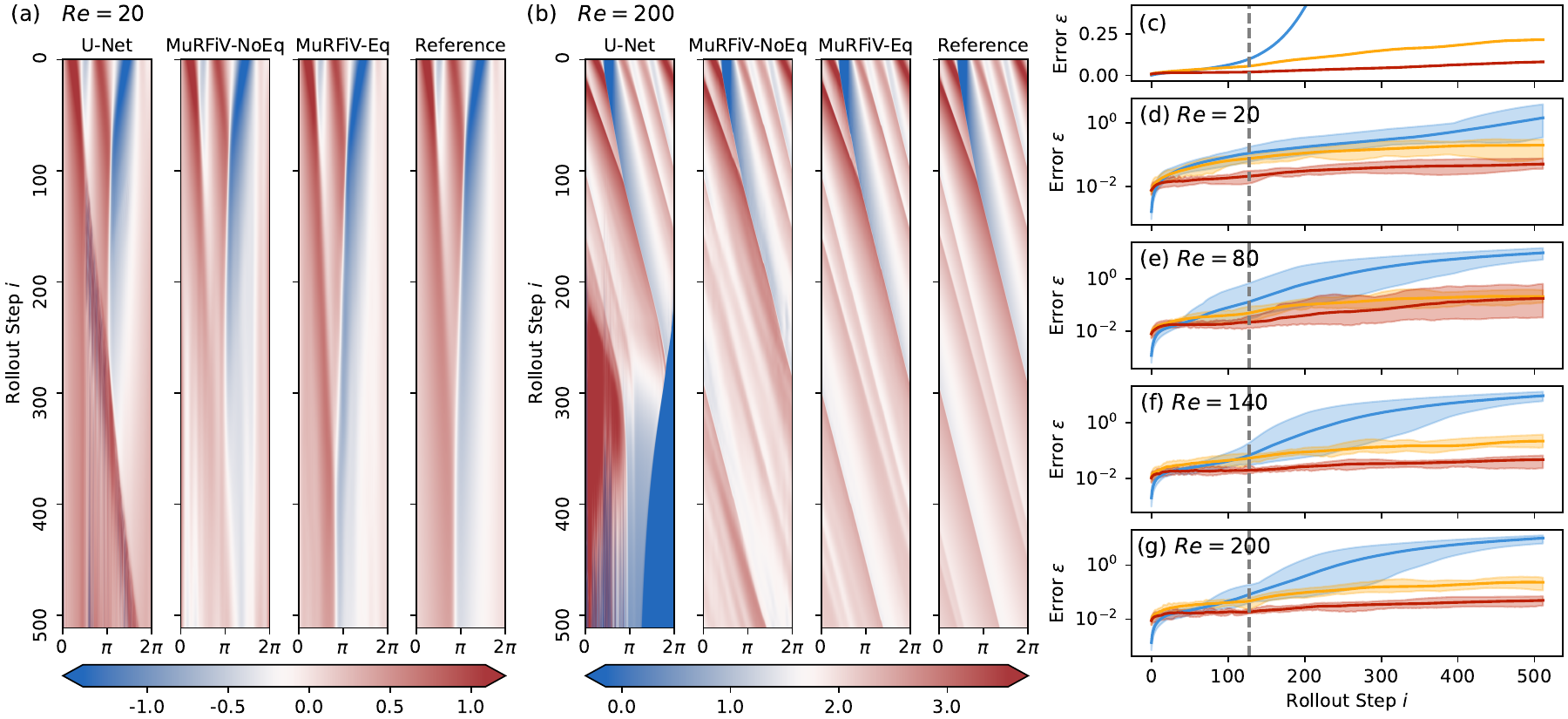}
    \caption{Future forecasting with training Reynolds number and starting from training initial conditions in the one-dimensional Burgers' Equation case. The first $128$ time steps are used for training, while the following $384$ time steps are considered ``extrapolation'' in time. (a) and (b), contours of the prediction results made by different models at the lowest and highest Reynolds numbers in the training set, respectively. (c-g) The mean prediction error $\varepsilon_i$ averaged over all the cases (c) and the error averaged over specific Reynolds numbers (d-g) are represented by the solid lines. The blue curve indicates \unet{} (\mkline{myblue}), while \mrf{} without equation (\mkline{myyellow}) and \mrf{} with equation (\mkline{myred}) are represented by yellow and red curves, respectively. The gray dashed lines (\mkline[dashed]{gray}) indicate the end of training time steps. The shaded areas in (d-g) represent the envelope of errors of all the test cases with the same Reynolds number.}
    \label{fig:b1d_train}
\end{figure}

\begin{figure}[!h]
    \centering
    \includegraphics[width=\linewidth]{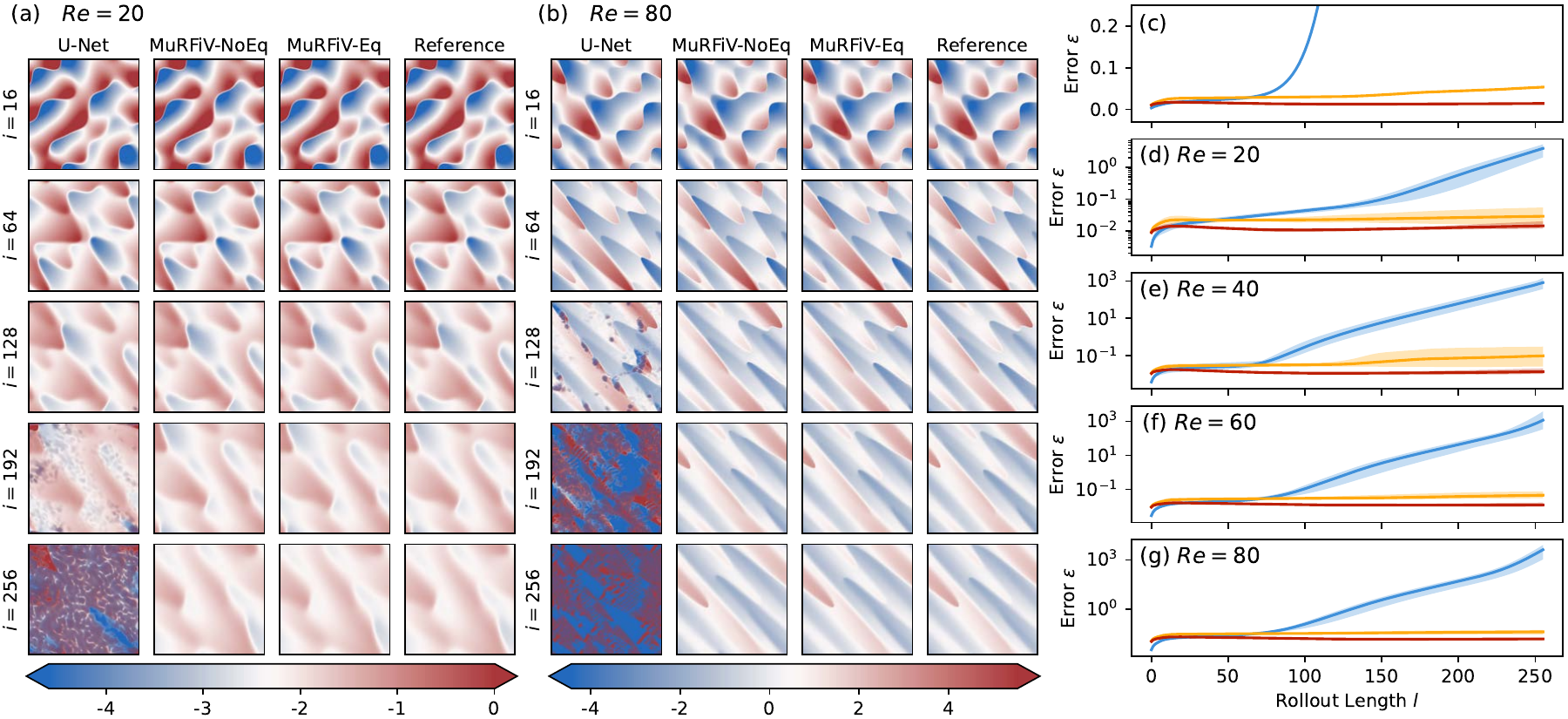}
    \caption{Forecasting results for the 2D Burgers' equation using training parameters. (a) and (b) Comparison of predicted scalar fields at the final time step for representative cases. (c-g) Time evolution of the mean prediction error averaged over the dataset (c) and for specific Reynolds numbers (d-g). Blue: \unet{} (\mkline{myblue}); Yellow: \mrf{}-NoEq (\mkline{myyellow}); Red: \mrf{}-Eq (\mkline{myred}).}
    \label{fig:b2d_Train}
\end{figure}

\begin{figure}[!h]
    \centering
    \includegraphics[width=\linewidth]{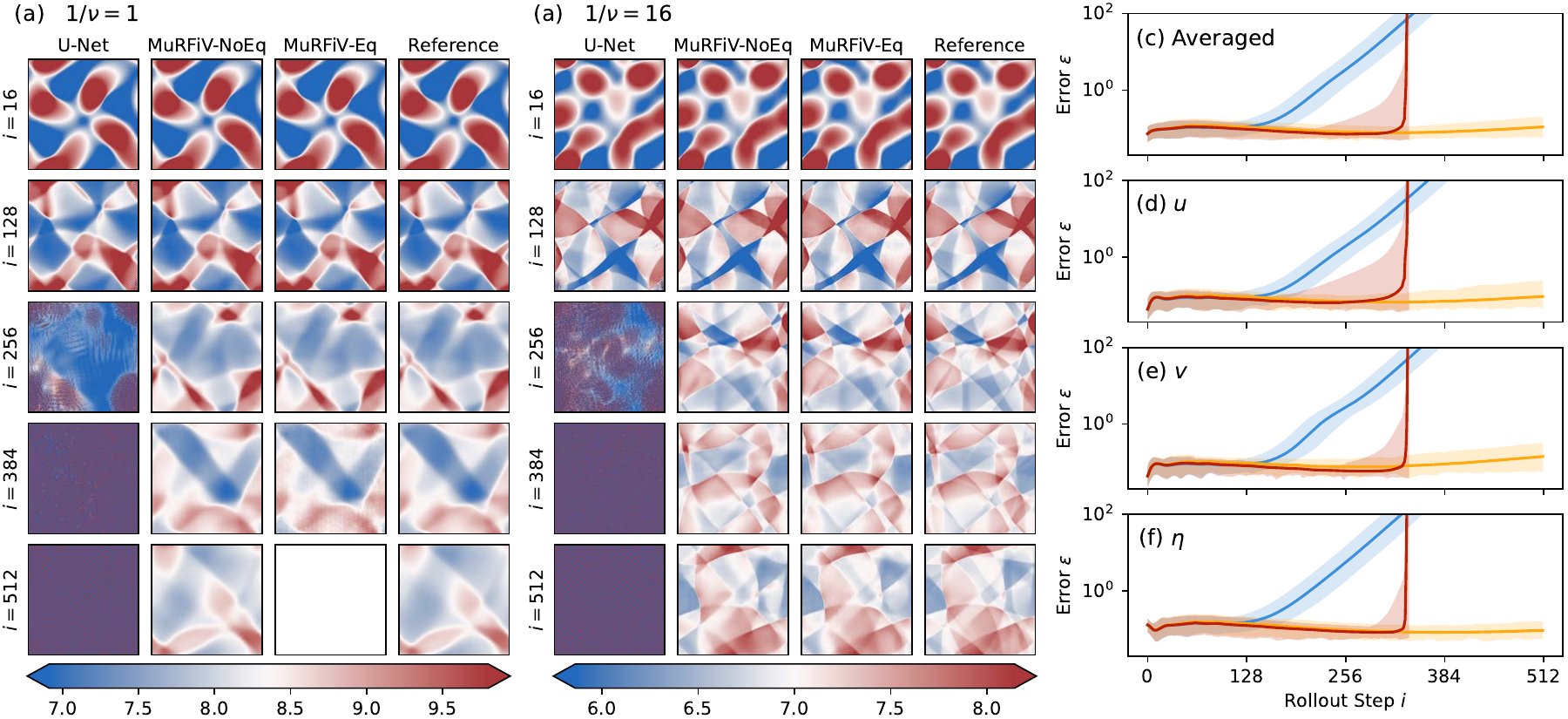}
    \caption{Forecasting results for the Shallow Water equations using training parameters. (a) and (b) Comparison of predicted fields at the final time step for representative cases. (c-g) Time evolution of the mean prediction error averaged over the dataset (c) and for specific viscosities (d-g). Blue: \unet{} (\mkline{myblue}); Yellow: \mrf{}-NoEq (\mkline{myyellow}); Red: \mrf{}-Eq (\mkline{myred}).}
    \label{fig:sw_train}
\end{figure}

\begin{figure}[!h]
    \centering
    \includegraphics[width=\linewidth]{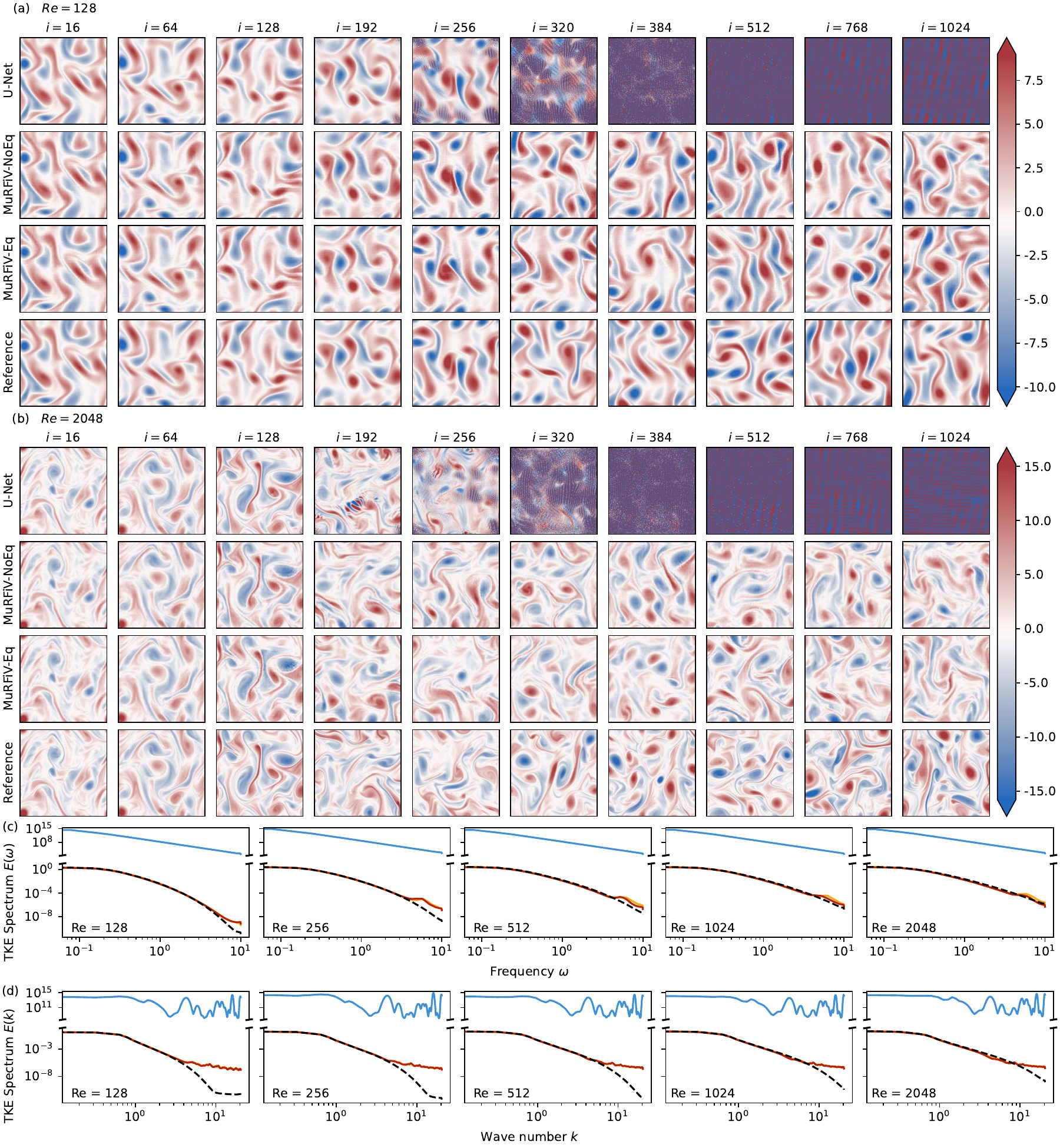}
    \caption{Forecasting results for Kolmogorov flow using training parameters. (a) and (b) Comparison of instantaneous vorticity fields at the final time step. (c) and (d) Turbulent Kinetic Energy (TKE) spectrum averaged over the evaluation window. Blue: \unet{} (\mkline{myblue}); Yellow: \mrf{}-NoEq (\mkline{myyellow}); Red: \mrf{}-Eq (\mkline{myred}).}
    \label{fig:ns_train}
\end{figure}

\end{document}